\documentclass[aps,prl,twocolumn,superscriptaddress,floatfix]{revtex4-1}
\usepackage{amssymb}
\usepackage{amsmath}
\usepackage{wasysym}
\usepackage{verbatim}
\usepackage[colorlinks,bookmarks=false,citecolor=blue,linkcolor=red,urlcolor=blue]{hyperref}
\usepackage[usenames,dvipsnames]{color}
\usepackage{times}
\usepackage{color}
\usepackage{graphicx}
\usepackage{epsfig}
\usepackage{float}
\usepackage{bbold}
\usepackage{tabularx}
\usepackage{natbib}
\usepackage{pdfpages}

\pdfoutput=1
\newcommand {\ket}[1] {|#1 \rangle}
\newcommand {\bra}[1] {\langle#1 |}

\newcommand{\Eref}[1]{Eq. (\ref{#1})}

\newcommand{\Fref}[1]{Fig. \ref{#1}}

\newcommand{\kB}{k_\mathrm{B}}

\newcommand{\trace}{\mathrm{Tr}}
\newcommand{\be}{\begin{equation}}
\newcommand{\ee}{\end{equation}}
\newcommand{\bea}{\begin{eqnarray}}
\newcommand{\eea}{\end{eqnarray}}

\begin{document}

\title{``Light-cone'' dynamics after quantum quenches in spin chains}
\author{Lars Bonnes}
\email{lars.bonnes@uibk.ac.at}
\affiliation{Institute for Theoretical Physics, University of Innsbruck, A-6020 Innsbruck, Austria.}

\author{Fabian H. L. Essler}
\affiliation{The Rudolf Peierls Centre for Theoretical Physics, Oxford University, Oxford OX1 3NP, UK}

\author{Andreas M. L\"auchli}
\affiliation{Institute for Theoretical Physics, University of Innsbruck, A-6020 Innsbruck, Austria.}

\date{\today}

\begin{abstract}
Signal propagation in the non equilibirum evolution after quantum
quenches has recently attracted much experimental and theoretical interest.
A key question arising in this context is what principles, and which of
the properties of the quench, determine the characteristic propagation
velocity. Here we investigate such issues for a class of quench
protocols in one of the central paradigms of interacting many-particle
quantum systems, the spin-1/2 Heisenberg XXZ chain. We consider
quenches from a variety of initial thermal density matrices to the
same final Hamiltonian using matrix product state methods. The
spreading velocities are observed to vary substantially with the
initial density matrix. However, we achieve a striking data collapse
when the spreading velocity is considered to be a function of 
the excess energy. Using the fact that the XXZ chain is integrable, 
we present an explanation of the observed velocities in terms of
``excitations'' in an appropriately defined  generalized Gibbs ensemble.
\end{abstract}

\maketitle 


The last few years have witnessed a number of significant advances in
understanding the nonequilibirum dynamics in isolated quantum systems.
Much of this activity has focussed on fundamental concepts such as
thermalization~\cite{rigol08,deutsch91,srednicki94,srednicki96,srednicki99}
or the roles played by
dimensionality and conservation laws~\cite{rigol07,iucci09,calabrese12,barthel08,cramer08,cramer10,fioretto10,calabrese11,calabrese12a,caux12,collura13}.

Another key issue concerns the spreading of correlations out of
equilibrium, and in particular the ``light-cone'' effect after global
quantum quenches. The most commonly studied protocol in this context
is to prepare the system in the ground state of a given Hamiltonian,
and to then suddenly change a system parameter such as a magnetic
field or interaction strength. At subsequent times the spreading of
correlations can then be analyzed by considering the time-dependence
of two-point functions of local operators separated by a fixed distance.
As shown by Lieb and Robinson~\cite{lieb72,sims10}, the
velocity of information transfer in quantum systems is bounded. This
gives rise to a causal structure in commutators of local operators at different times, although
Schr\"odinger's equation, unlike relativistic theories, has no built-in
speed limit. Recently, the Lieb-Robinson bounds have been
refined~\cite{juneman13,lieb13,kliesch13} and extended to mixed
state dynamics in open quantum systems~\cite{poulin10,kliesch13} as
well as creation of topological quantum order~\cite{bravyi06}. 

A striking consequence of the Lieb-Robinson bound is that the equal-time correlators after a quantum quench feature a ``light-cone'' effect~\cite{bravyi06}, which is most pronounced
for quenches to conformal field theories from initial density matrices
with a finite correlation length~\cite{calabrese06}:
connected correlations are initially absent, but exhibit a marked
increase after a time $t_0=x/2v$. This observation  
is explained by noting~\cite{calabrese05,calabrese07} that entangled
pairs of quasi-particles initially located half-way between the two
points of measurement, propagate with the speed of light $v$ and hence
induce correlations after a time $t_0$. These predictions have been
verified numerically in several systems, see
e.g.~\cite{dechiara06,laeuchli08,manmana09,hauke13,eisert13,carleo14}.  
Very recently light-cone effects after quantum quenches have been
observed in systems of ultra-cold atomic
gases~\cite{cheneau12,langen13} and trapped ions~\cite{jurcevic14,richerme14}.
The experimental work raises the poignant theoretical issue of
which velocity underlies the observed light-cone effect in
non-relativistic systems at finite energy densities. Here there is no
unique velocity of light, and quasi-particles in interacting systems
will generally have finite life times depending on the details of the
initial density matrix.

\begin{figure}
\includegraphics[width=\columnwidth]{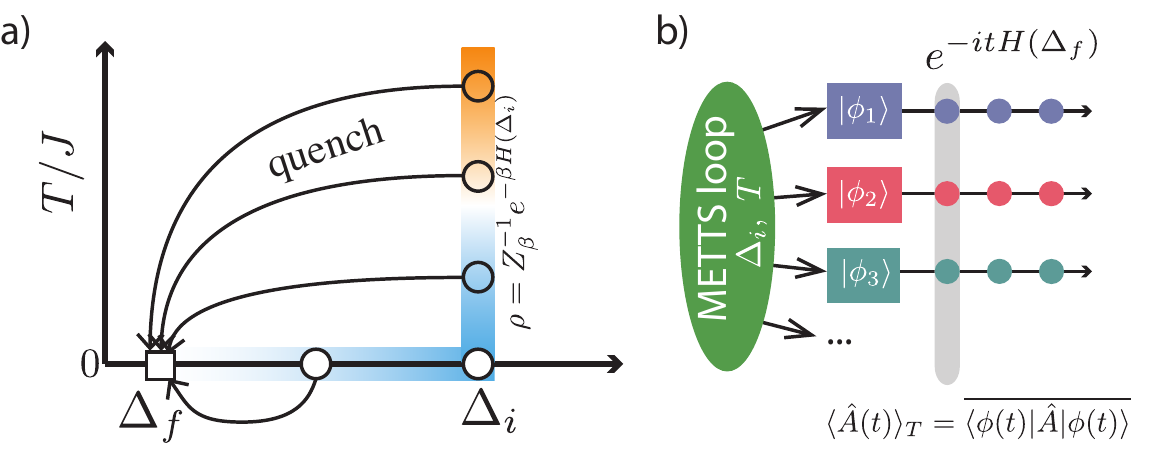}
\caption{(Color online) 
a) Quench protocol: The system is initially prepared in either the ground state of some Hamiltonian $H(\Delta_i)$ or in a thermal state $\rho=Z_\beta^{-1}\exp[-\beta H(\Delta_i)]$ with temperature $T=1/(\kB \beta)$. 
At time $0$, the anisotropy is quenched to $\Delta_f$ and we let the system evolve in time for various initial values of $\Delta_i$ and $T$.
b) Outline of the numerical procedure: The METTS projection loop generates generates an ensemble of wave function for some initial $\Delta_i$ and temperature $T$. Each realization is evolved in time and expectation values are obtained by averaging over the ensemble.
}
\label{fig:Protocol}
\end{figure}
 
In order to shed some light on this issue, we have carried out a
systematic study of the spreading of correlations in the spin-1/2
Heisenberg XXZ chain, a key paradigm among interacting many
body quantum systems in one spatial dimension.  We fix the final
(quenched) Hamiltonian and vary the initial conditions over a large
range of parameters. Moreover, we do not only consider initial pure
states~\cite{manmana09} but also prepare the system in thermal initial states as
illustrated in~\Fref{fig:Protocol}(a). The latter is of significant
interest in view of experimental realizations. 
Apart from a recent numerical study for local quenches~\cite{karrasch14}, 
the spreading of signals in quenches from thermal states is basically unexplored.

Our numerical simulations are based on a quench extension of a recently proposed 
algorithm utilizing an optimized wave function
ensemble called Minimally Entangled Typical Thermal States
(METTS)~\cite{white09,stoudenmire10} implemented within the matrix
product state (MPS) framework.  We come back to the description of the algorithm 
and a discussion of its performance towards the end of this paper.

\begin{figure}[t]
\includegraphics[width=0.9\columnwidth]{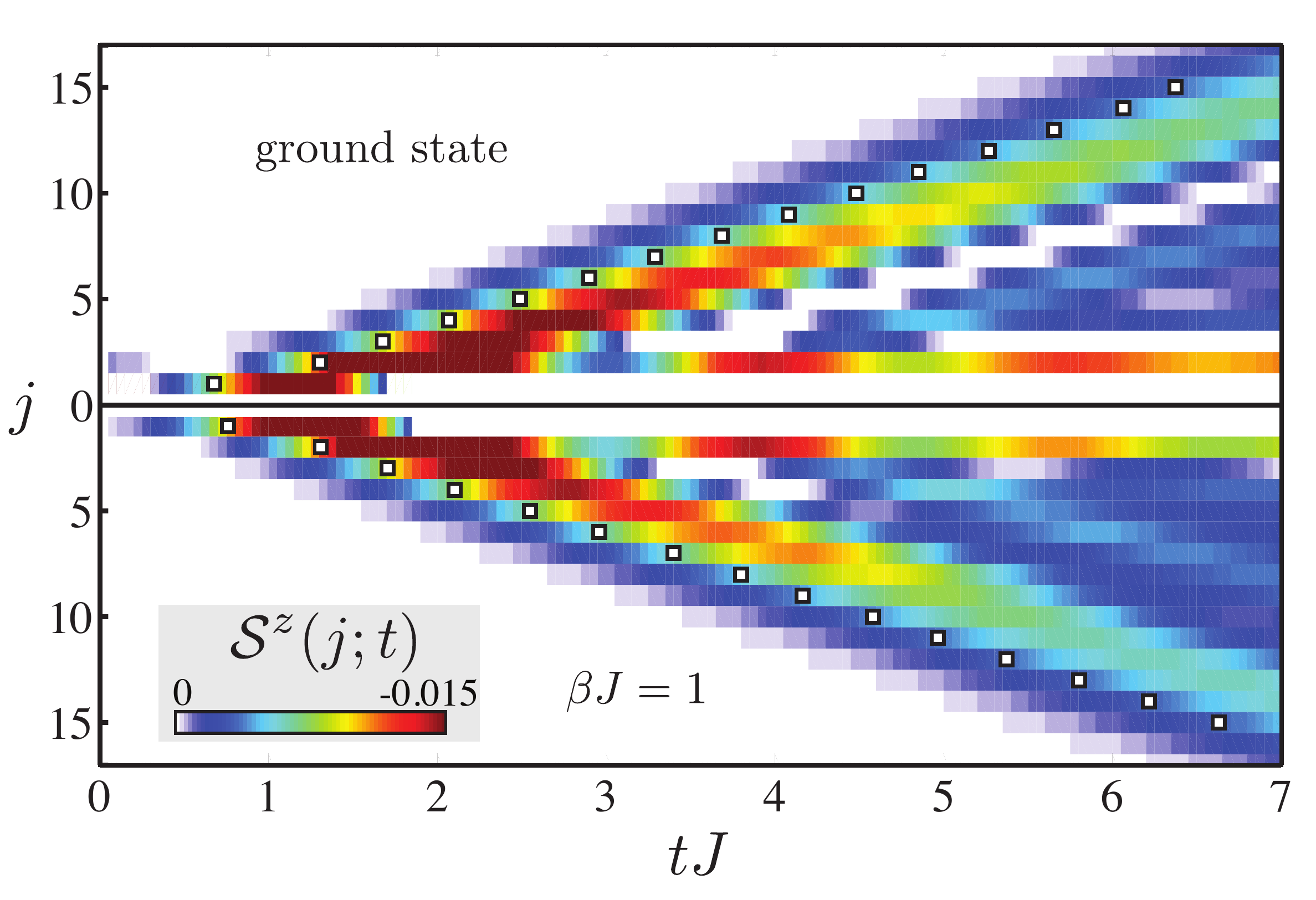}
\caption{  
Space-time plot of the $\mathcal{S}^z$ correlation functions \eqref{eqn:szcorrelator} for the quench from $\Delta_i=4$ to $\Delta_f=\cos(\pi/4)$.
This particular value of the final interaction is chosen due to technical reason in the Bethe ansatz calculations.
The upper panel shows ground state data whereas the lower panel shows data from a thermal density matrix at $T/J=1$. 
This illustrates that the light-cone effect in this observable persists also at finite temperatures.
}
\label{fig:fig2}
\end{figure}

\paragraph*{Results.---}
In the following we consider quenches to the spin-1/2 Heisenberg XXZ chain
with anisotropy $\Delta$
\begin{equation}
	H(\Delta)= J\sum_{i=1}^{L-1} \left(S_i^x S_{i+1}^x + S_i^y S_{i+1}^y + \Delta S_i^z S_{i+1}^z \right).
\end{equation}
Initially, the system is prepared in a Gibbs state corresponding to an
XXZ Hamiltonian with anisotropy $\Delta_i$ at a temperature $T$, i.e.
\be
\rho(t=0)=Z_\beta^{-1}\exp[-\beta H(\Delta_i)]\ , \quad \beta=\frac{1}{\kB T},
\ee
where $Z_\beta=\trace \exp[-\beta H(\Delta_i)]$ (we set $\kB=1$).
The anisotropy is then quenched at time $t=0^+$ from $\Delta_i$ to
$0\leq\Delta_f\leq 1$, as depicted in \Fref{fig:Protocol}(a), and the system
subsequently evolves unitarily with Hamiltonian
$H(\Delta_f)$~\footnote{Note that the system is \textit{not} connected
to a heat bath during the time evolution and energy $\trace[H \rho(t)]$
is conserved.}. In order to probe the spreading of correlations we consider
the longitudinal spin correlation functions
\begin{equation}
\mathcal{S}^z(j;t) = \langle S_{L/2}^z(t) S_j^z(t) \rangle - \langle S_{L/2}^z(t)\rangle \langle S_j^z(t)\rangle
\label{eqn:szcorrelator}
\end{equation}
centered around the middle of the chain. Results for
$\mathcal{S}^z(j;t)$ are most easily visualized in space-time plots,
and typical results are shown in \Fref{fig:fig2}. The most
striking feature observed in these plots is the light-cone effect:
at a given separation $j$ connected correlations $\mathcal{S}^z(j;t)$
arise fairly suddenly at a time that scales linearly with $j$.

\begin{figure}[t]
\includegraphics[width=0.9\columnwidth]{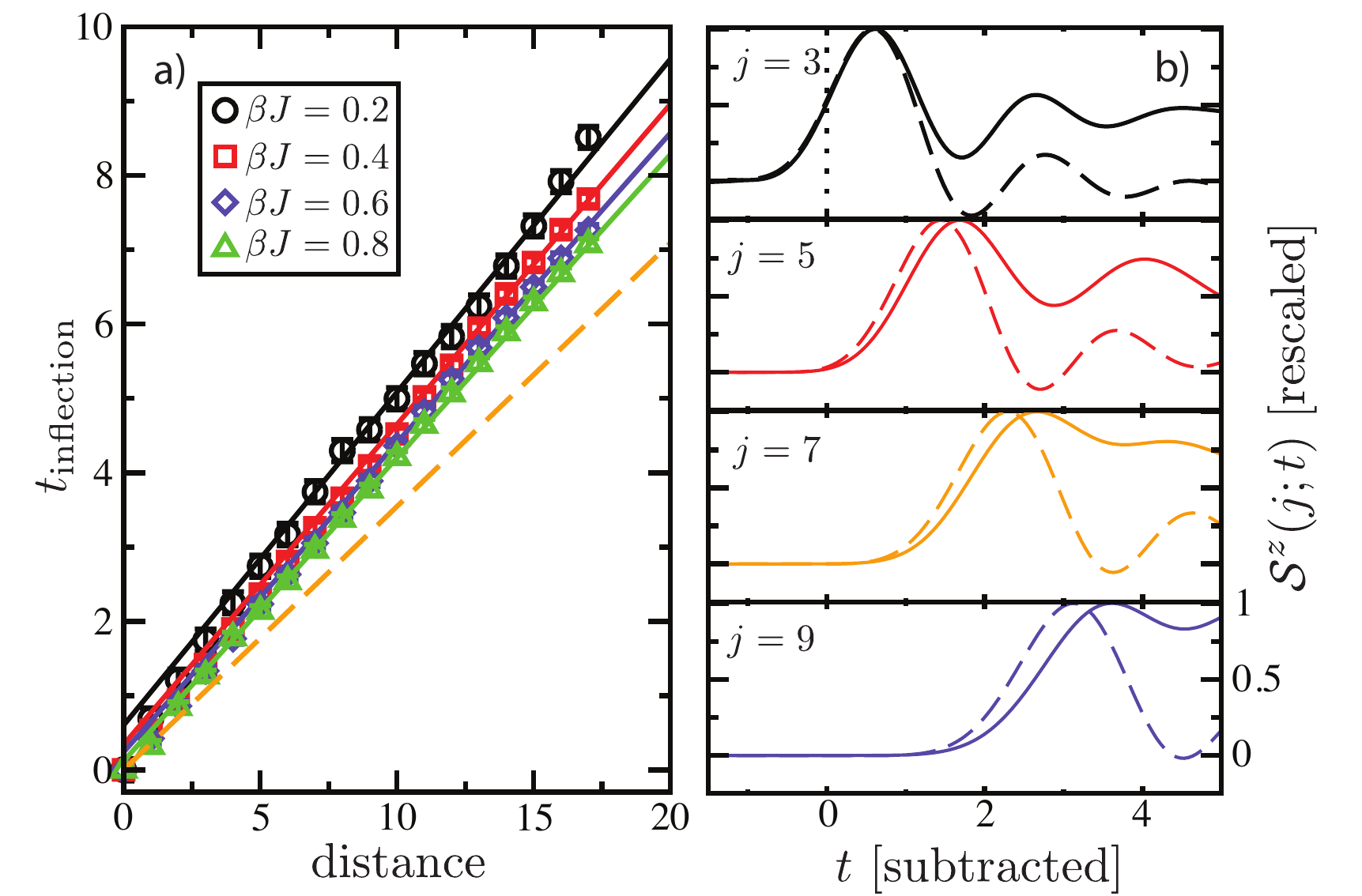}
\caption{  
\textit{a)} Extracted inflection points versus distance for different initial temperatures for the quench from $\Delta=4$ to $\cos(\pi/4)$.
The straight lines correspond to the velocities extracted from the GGE where only the offset of the time axis has been fitted. The orange dashed line 
denotes the ground state Bethe ansatz velocity at $\Delta_f$.
\textit{b)} Rescaled averaged spin correlation functions for the quench from $\Delta=4$ to $\cos(\pi/4)$ for $T/J=1$ and the ground state (dashed line) and different distances $j=3$, 5, 7 and 9.
We omit the error bars for clarity of the figure.
The time axis is relative to the first inflection point of the correlation functions for $j=3$.
One can see that the signal is delayed as the initial temperature is increased.
}
\label{fig:fig3}
\end{figure}

These results demonstrate that the light-cone effect persists for
mixed initial states, although the visibility of the signal is 
diminished with increasing temperature (until it vanished
completely at $\beta=0$ since the initial density matrix is trivial and stationary). 
Comparing the time evolution of the correlation functions for
different initial temperatures, we see (cf \Fref{fig:fig2} and \Fref{fig:fig3})
that the signal front is delayed when the temperature of the initial
state is increased, signalling that the spreading slows down. We
further observe that the spreading velocity is sensitive to the
strength of the quench, i.e. the value of the initial interaction.
At this point we should note that this finding is unexpected. Based on 
our current understanding of quenches to CFTs or of Lieb-Robinson
bounds, there are no predictions available which support
spreading velocities depending on the initial state.

Having established the result that the spreading velocity
depends both on the initial density matrices and the final
Hamiltonian, an obvious question is which properties of $\rho(t=0)$  
are relevant in this context. 
In order to quantify this aspect we define the precise location of the
light-cone as the first inflection point of the signal front
observed in $\mathcal{S}^z$ (alike Ref.~\onlinecite{manmana09}). This allows us to extract a spreading
velocity $v_s$ by performing a linear fit to the largest accessible
time, where expected finite-distance effects~\cite{barmettler12} are
small. 

\begin{figure}[t]
\includegraphics[width=\columnwidth]{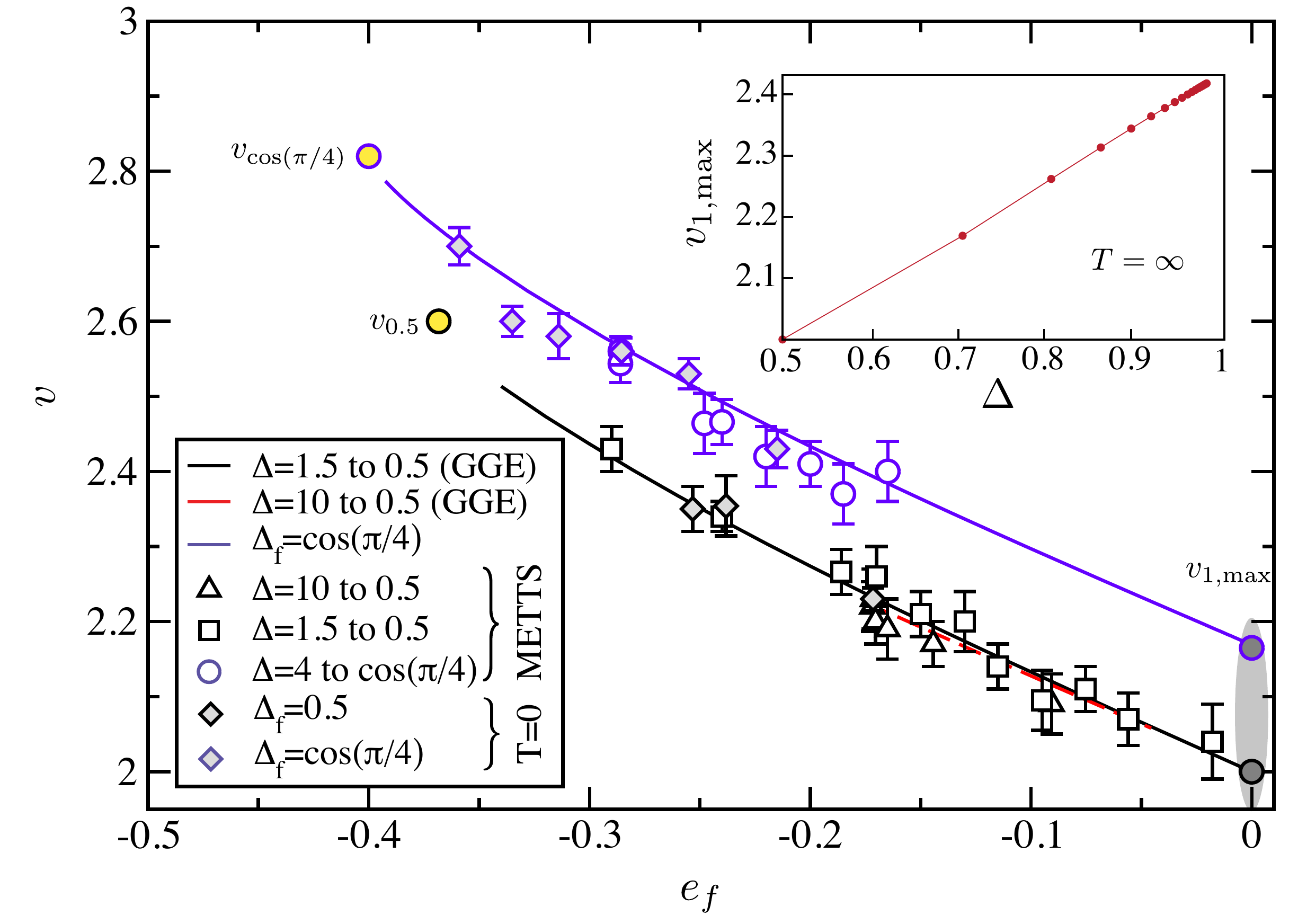}
\caption{ (Color online)
Spreading velocity $v_s$ extracted from the spin correlation function
$\mathcal{S}^z$ as a function of the final energy $e_f$ density for $\Delta_f
= 1/2$ and  $\cos \pi /4$. The symbols denote numerical results
obtained from either thermal or pure initial states with different
$\Delta_i$. 
The blue and black solid lines denote the spreading velocities from TBA using only the energy density
wheres the red line shows the results for the quench from $\Delta_i=10$ to $0.5$ using also the first conserved quantity.
The corresponding velocities for the quench from $\Delta_i=1.5$ lie on top of the black line, i.e. the GGE effects are smaller than the line width.
The rightmost symbols denote $v_{\Delta_f}$ at the energy density of the ground state whereas the right most ones denote $v_{1,\mathrm{max}}$.
The inset shows
the velocity at $\beta=0$, $v_{1,\mathrm{max}}$, extracted from the thermodynamic Bethe
ansatz for $\Delta = \cos(\pi/n)$. 
}
\label{fig:fig4}
\end{figure}

Our main result, shown in
\Fref{fig:fig4}, is that the spreading velocity is mainly
determined by the \emph{final energy density}
\be
e_f =\frac{\trace[H(\Delta_f) \rho(t=0)]}{L}.
\ee
Plotting the measured velocities against $e_f$ leads to a remarkable data
collapse for a variety of quenches from thermal as well as pure initial
states for various $\Delta_i$. This holds in spite of the fact that
the system is integrable and thus its dynamics is constrained by an
infinite set of conserved quantities. As we will show in the
following, the observed velocities can be explained quantitatively by
considering ``excitations'' in an appropriately defined generalized Gibbs
ensemble.   

Focusing on the quenches to $\Delta_f=1/2$ as well as $\cos(\pi/4)\approx0.707$, we observe that the
spreading velocity $v_s$ decreases significantly as the final energy density
is increased by increasing $T$ or altering $\Delta_i$. The numerical data suggests that $v_s$
approaches a non-trivial velocity in the infinite-temperature limit that depends on 
$\Delta_f$. In fact this velocity can be obtained from Bethe ansatz (see discussion below)
and is shown for the series $\Delta_f=\cos(\pi/n)$ in the inset of \Fref{fig:fig4}. 
For very weak quenches, where only the low-energy (relative to the ground
state of $H(\Delta_f)$) degrees of freedom become populated, one
expects that the spreading velocity is given by the maximal mode
velocity $v_\Delta=\pi [(1-\Delta^2)/(2
  \arccos{\Delta})]^{-1/2}$.   
In fact, the spreading velocity extrapolates to $v_\mathrm{\Delta_f}$,
when the final energy approaches the ground state energy of
$H(\Delta_f)$. For the non-interacting case $\Delta_f=0$ which reduces
essentially to free fermions, we find that the spreading velocity for
all initial conditions is compatible with the maximal mode velocity,
$v_0=2$. This is consistent with results we obtained for quenches to
the critical point of a one-dimensional Ising model in a transverse
field, which is essentially also a free theory, where also no
significant dependence of the spreading velocity on the initial
conditions was observed.  

We now provide a theoretical explanation of our striking numerical observations.

\paragraph*{Excitations in a Generalized Gibbs Ensemble.---} 
A recent work ~\cite{caux13} proposed that correlation functions of local
operators after a quench to an integrable model, prepared in a pure
state $|\Psi\rangle$, are given by
\begin{eqnarray}
\lim_{L \rightarrow \infty} \langle {\cal O}(t) \rangle 
= \lim_{L \rightarrow \infty} \left[
\frac{\langle \Psi | {\cal O}(t) | \Phi_{s} \rangle}{2\langle \Psi
  | \Phi_{s} \rangle}
+\Phi_s\leftrightarrow\Psi\right].
\label{eq:Ot1a}
\end{eqnarray}
Here $|\Phi\rangle_s$ is a simultaneous eigenstate of the post-quench
Hamiltonian and all local, higher conservation laws $I_n$, such that
\bea
i_n\equiv
\lim_{L\to\infty}\frac{1}{L}\trace[\rho(t=0)I_n]
=
\lim_{L\to\infty}\frac{1}{L}\frac{\langle\Phi_s|I_n|\Phi_s\rangle}{\langle\Phi_s|\Phi_s\rangle}.
\eea
In the case of interest here we have ${\cal O}
(t)=S^z_{L/2}(t)S^z_j(t)$. Importantly, the state $|\Phi_s\rangle$ can
be constructed by means of a generalized Thermodynamic Bethe Ansatz
(gTBA)~\cite{mossel12,demler12}. The stationary state itself is
expected to be described by an appropriate GGE involving the known
ultra-local~\cite{grabowski95} and quasi-local~\cite{prosen13}
conservation laws, and possibly others\cite{pozsgay13,fagotti13,fagotti14,wouters14,poszgay14}.  

It was argued in Ref.~\onlinecite{caux13} that states obtained by making
microscopic changes to $|\Phi_s\rangle$ are most important to describe
the dynamics at (sufficiently) late times. This is motivated by
employing a Lehmann representation in terms of energy eigenstates
$H(\Delta_f)|n\rangle=E_n|n\rangle$ 
\bea
\langle \Psi | {\cal O}(t) | \Phi_{s} \rangle
=\sum_n\langle \Psi |n\rangle\langle n| {\cal O} | \Phi_{s} \rangle
e^{-i(E_n-E_{\Phi_s})t},
\eea
and noting that at sufficiently late times only states with
$(E_n-E_{\Phi_s})/J={\cal O}(1)$ are likely to contribute due to the
otherwise rapidly oscillating phase.
It is then tempting to conjecture that spreading of correlations
occurs through these ``excited states'' (which by constructed can have
either positive or negative energies relative to the representative
state), and the light-cone effect propagates with the maximum group
velocity that occurs amongst them. The method for calculating such
excited state velocities is depicted schematically in
Fig.~\ref{fig:illustration}, and details of the calculations are
provided in the Supplementary Material. The basic idea is to use TBA
methods to determine the macrostate minimizing the generalized Gibbs
free energy. This is characterized by appropriate particle/hole
distribution functions $\rho^{p,h}_j(x)$ for of elementary excitations
labelled by the index $j$ ($x$ parametrizes the respective
momenta). The corresponding ``microcanonical''
description~\cite{pozsgay11,cassidy11,caux13} is based on the 
particular simultaneous eigenstate $|\Phi_s\rangle$ of the Hamiltonian
and the higher conservation laws, characterized by the set
$\{\rho^{p,h}_j(x)\}$ in the thermodynamic limit. One then considers 
small changes of this microstate, and determines the resulting ${\cal
  O}(1)$ (i.e. non-extensive) changes in energy and momentum. These
can be described in terms of additive ``elementary excitations''
relative to $|\Phi_s\rangle$. Finally, one determines the dispersion
relations and hence the group velocities of these excitations. 
\begin{figure}[t]
\includegraphics[width=0.7\columnwidth]{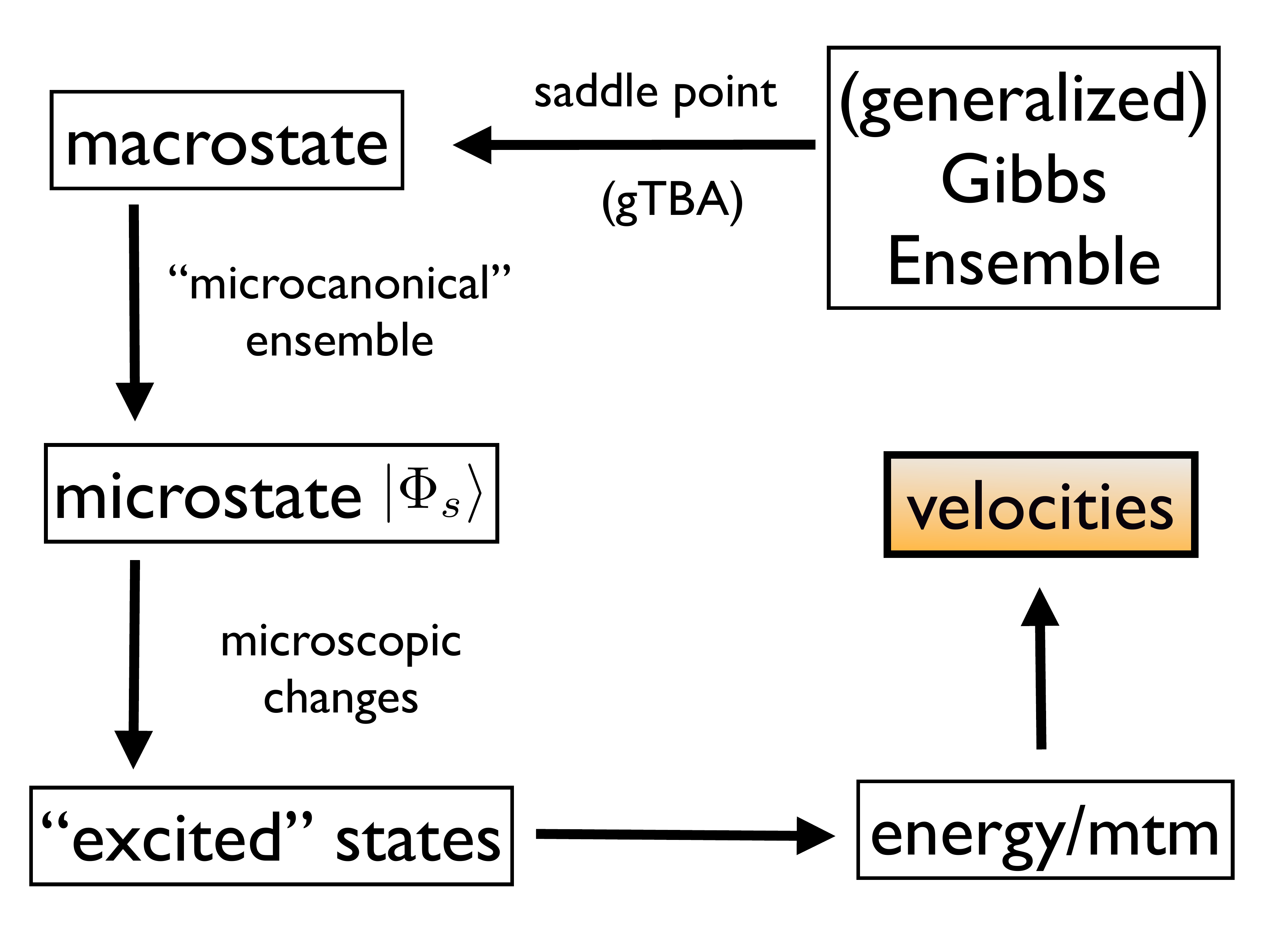}
\caption{ 
Scheme for the extraction of the velocities from the GGE.
See text for details.
}
\label{fig:illustration}
\end{figure}
The most significant qualitative features of the ``GGE
excitation spectrum'' obtained in this way are as follows. 
(i) There are several types of infinitely long-lived elementary
excitations. (ii) Their number depends only on the anisotropy
$\Delta_f$ 
~\cite{takahashi72}, but their dispersions are sensitive to the full set
$\{i_n\}$ of (conserved) expectation values. (iii) In practice, we
need to compute $i_n$ numerically. Given that explicit expressions for
$I_n$ become rapidly extremely complicated 
~\cite{grabowski95}, we retain only two
conservation  laws, namely energy and $I_3$, which involves 4-spin
interactions ($I_2$ is odd under time reversal
and hence does not play a role for the quenches considered here). This
can be justified by noting that the differences in the calculated
maximal velocities between a Gibbs ensemble and a GGE with one added
conservation law are small, and that the most local conservation laws
are most important for accurately describing the properties of local
operators~\cite{fagotti13b}.
(iv) In the cases we have considered, the maximal propagation velocity
is found for the same type of excitation (``positive parity 1-strings''\cite{takahashi72}).
The results for the maximal velocities obtained from this gTBA
analysis are compared to our numerical computations in the main panel of 
\Fref{fig:fig4}. The agreement is clearly very good.
Also, the inset if \Fref{fig:fig4} shows the velocities for infinite
temperatures ($e_f=0$) for $\Delta_f=\cos(\pi/n)$, where $n$ is an
integer, revealing a non-trivial $\Delta_f$ dependence even in this
limiting case. 

\paragraph*{Numerical method.---}
After having provided the physical results we shortly review the numerical procedure employed to simulate the mixed state dynamics.
MPS provide a powerful framework to study the real-time dynamics of one-dimensional quantum systems.
Originally conceived for ground state calculations~\cite{vidal03}, extensions to finite temperatures include purification schemes~\cite{feiguin05b,barthel09,karrasch12,barthel13,karrasch13b}, superoperators/matrix product operators (MPOs)~\cite{zwolak04,verstraete04b,prosen10} or transfer matrices~\cite{bursill96,wang97,sirker05}.
Very recently Refs.~\cite{white09,stoudenmire10} introduced a stochastic method in which the expectation value of a thermal density matrix is replaced by an average over an ensemble of wave functions, $\lbrace \ket{\phi_i} \rbrace$, that \textit{i)} can be efficiently sampled (importance sampling) using Markov chains and 
\textit{ii)} only hosts the minimal (small) amount of entanglement required at that temperature, thus allowing for an efficient representation in terms of MPS.
This ensemble was therefore called METTS (\underline{m}inimally \underline{e}ntangled \underline{t}ypical \underline{t}hermal \underline{s}tates)~\cite{white09,stoudenmire10}. 

We show that the METTS method, thus far only applied to static equilibrium problems, can be easily extended to study real time evolution by realizing that the expectation 
value of some real-time propagated operator $\hat A(t)$ can be written as 
\begin{equation}
	\langle \hat A(t) \rangle_T = \frac{1}{Z_\beta}\trace e^{-\beta H} \hat{A}(t) 
			= \overline{\bra{\phi_i(t)} \hat A \ket{\phi_i(t)}}
	\label{eq:exvalue}
\end{equation}
where the last term denotes an average over the time-evolved METTS ensemble~\footnote{It is possible to
evaluate observables that commute with the projection operator using
the CPS rather than the METTS. The real time evolution, though,
prohibits us from taking advantage of this improved measurement
scheme.}.
We employ the numerical scheme illustrated in \Fref{fig:Protocol}b) where we first generate an ensemble of wave functions following Ref.~\onlinecite{stoudenmire10}.
In a second step each $\Phi_i$ is evolved in time using the TEBD algorithm~\cite{vidal03} and \Eref{eq:exvalue} is evaluated.
We average over a few hundred METTS instances and are limited due to runaway phenomena~\cite{gobert05} to times of $tJ \sim 6 - 8$.
Due to the reachable time scales we consider systems sizes of up to 50 sites here, but studying larger systems poses no particular problem by itself.
A detailed description of the numerical method used here can be found in the Supplementary Material.

Compared to a complementary approach, where the von Neumann equation for the full system density matrix is integrated within an matrix product operator framework~\cite{zwolak04}, 
we find that the METTS approach is able to reach significantly longer times, and we therefore believe that the METTS approach is quite promising to study global quenches at finite temperature.
A full comparison of the different approaches, however, is beyond the scope of this paper and will be address in a forthcoming publication~\cite{bonnes14}.

\paragraph*{Conclusions.---}
We have analyzed the spreading of correlations after quantum quenches
in the spin-1/2 Heisenberg XXZ chain. Our initial density matrix
describing the system was taken to be a Gibbs distribution at a
particular temperature and initial value of anisotropy
$\Delta_i$. We observed a pronounced light-cone effect in the
connected longitudinal spin-spin correlation function. We found that
the propagation velocity $v$ of the light-cone depends not only on the
final Hamiltonian, but also on the initial density matrix. For the
quenches we considered the observed values of $v$ are
well-characterized by the expectation value of the final Hamiltonian
in the initial state. These findings were found to be in accord with
expectation based on properties of ``excitations'' in an appropriately
defined generalized Gibbs ensemble. We also have shown that one can
apply the METTS framework to study dynamical properties using
MPS. Although the method also exhibits the typical runaway behavior,
the lack of ancillary degrees of freedom or enlarged local Hilbert
spaces reduces the complexity of the simulations and a direct
comparison to other methods will be provided in a separate
publication~\cite{bonnes14}.  

Our work raises a number of interesting issues. First, we expect that
a full characterization of $v$ will involve not only the final energy
density, but the densities of all final higher conservation laws as
well. In fact, we observe that the effects of higher conserved
quantities are much more pronounced for negative $\Delta_f$ and this
point is under investigation. 
Second, our work raises the question, whether 
horizon effects related to slower excitations can become visible for
particular initial density matrices (in the case of local
quenches this is indeed the case~\cite{ganahl12}). Finally, our work
suggests that light-cone propagation in generic non-integrable models
ought to be rather non-trivial and warrants detailed investigation.

\paragraph*{Acknowledgements.---}  
We thank S. R. Manmana for discussions and acknowledge R. Bartenstein 
for collaboration at an early stage of the project.
This work was supported by the Austrian Ministry of Science BMWF as
part of the UniInfrastrukturprogramm of the Forschungsplattform
Scientific Computing at LFU Innsbruck, by the FWF SFB FOQUS 
and by the EPSRC under grants EP/I032487/1 and EP/J014885/1.

\bibliographystyle{apsrev4-1}

\begin{thebibliography}{71}%
\makeatletter
\providecommand \@ifxundefined [1]{%
 \@ifx{#1\undefined}
}%
\providecommand \@ifnum [1]{%
 \ifnum #1\expandafter \@firstoftwo
 \else \expandafter \@secondoftwo
 \fi
}%
\providecommand \@ifx [1]{%
 \ifx #1\expandafter \@firstoftwo
 \else \expandafter \@secondoftwo
 \fi
}%
\providecommand \natexlab [1]{#1}%
\providecommand \enquote  [1]{``#1''}%
\providecommand \bibnamefont  [1]{#1}%
\providecommand \bibfnamefont [1]{#1}%
\providecommand \citenamefont [1]{#1}%
\providecommand \href@noop [0]{\@secondoftwo}%
\providecommand \href [0]{\begingroup \@sanitize@url \@href}%
\providecommand \@href[1]{\@@startlink{#1}\@@href}%
\providecommand \@@href[1]{\endgroup#1\@@endlink}%
\providecommand \@sanitize@url [0]{\catcode `\\12\catcode `\$12\catcode
  `\&12\catcode `\#12\catcode `\^12\catcode `\_12\catcode `\%12\relax}%
\providecommand \@@startlink[1]{}%
\providecommand \@@endlink[0]{}%
\providecommand \url  [0]{\begingroup\@sanitize@url \@url }%
\providecommand \@url [1]{\endgroup\@href {#1}{\urlprefix }}%
\providecommand \urlprefix  [0]{URL }%
\providecommand \Eprint [0]{\href }%
\providecommand \doibase [0]{http://dx.doi.org/}%
\providecommand \selectlanguage [0]{\@gobble}%
\providecommand \bibinfo  [0]{\@secondoftwo}%
\providecommand \bibfield  [0]{\@secondoftwo}%
\providecommand \translation [1]{[#1]}%
\providecommand \BibitemOpen [0]{}%
\providecommand \bibitemStop [0]{}%
\providecommand \bibitemNoStop [0]{.\EOS\space}%
\providecommand \EOS [0]{\spacefactor3000\relax}%
\providecommand \BibitemShut  [1]{\csname bibitem#1\endcsname}%
\let\auto@bib@innerbib\@empty
\bibitem [{\citenamefont {Rigol}\ \emph {et~al.}(2008)\citenamefont {Rigol},
  \citenamefont {Dunjko},\ and\ \citenamefont {Olshanii}}]{rigol08}%
  \BibitemOpen
  \bibfield  {author} {\bibinfo {author} {\bibfnamefont {M.}~\bibnamefont
  {Rigol}}, \bibinfo {author} {\bibfnamefont {V.}~\bibnamefont {Dunjko}}, \
  and\ \bibinfo {author} {\bibfnamefont {M.}~\bibnamefont {Olshanii}},\ }\href
  {\doibase doi:10.1038/nature06838} {\bibfield  {journal} {\bibinfo  {journal}
  {Nature}\ }\textbf {\bibinfo {volume} {452}},\ \bibinfo {pages} {854}
  (\bibinfo {year} {2008})}\BibitemShut {NoStop}%
\bibitem [{\citenamefont {Deutsch}(1991)}]{deutsch91}%
  \BibitemOpen
  \bibfield  {author} {\bibinfo {author} {\bibfnamefont {J.~M.}\ \bibnamefont
  {Deutsch}},\ }\href {\doibase 10.1103/PhysRevA.43.2046} {\bibfield  {journal}
  {\bibinfo  {journal} {Phys. Rev. A}\ }\textbf {\bibinfo {volume} {43}},\
  \bibinfo {pages} {2046} (\bibinfo {year} {1991})}\BibitemShut {NoStop}%
\bibitem [{\citenamefont {Srednicki}(1994)}]{srednicki94}%
  \BibitemOpen
  \bibfield  {author} {\bibinfo {author} {\bibfnamefont {M.}~\bibnamefont
  {Srednicki}},\ }\href {\doibase 10.1103/PhysRevE.50.888} {\bibfield
  {journal} {\bibinfo  {journal} {Phys. Rev. E}\ }\textbf {\bibinfo {volume}
  {50}},\ \bibinfo {pages} {888} (\bibinfo {year} {1994})}\BibitemShut
  {NoStop}%
\bibitem [{\citenamefont {Srednicki}(1996)}]{srednicki96}%
  \BibitemOpen
  \bibfield  {author} {\bibinfo {author} {\bibfnamefont {M.}~\bibnamefont
  {Srednicki}},\ }\href {\doibase 10.1088/0305-4470/29/4/003} {\bibfield
  {journal} {\bibinfo  {journal} {J. Phys. A}\ }\textbf {\bibinfo {volume}
  {29}},\ \bibinfo {pages} {L75} (\bibinfo {year} {1996})}\BibitemShut
  {NoStop}%
\bibitem [{\citenamefont {Srednicki}(1999)}]{srednicki99}%
  \BibitemOpen
  \bibfield  {author} {\bibinfo {author} {\bibfnamefont {M.}~\bibnamefont
  {Srednicki}},\ }\href {\doibase 10.1088/0305-4470/32/7/007} {\bibfield
  {journal} {\bibinfo  {journal} {J. Phys. A}\ }\textbf {\bibinfo {volume}
  {32}},\ \bibinfo {pages} {1163} (\bibinfo {year} {1999})}\BibitemShut
  {NoStop}%
\bibitem [{\citenamefont {Rigol}\ \emph {et~al.}(2007)\citenamefont {Rigol},
  \citenamefont {Dunjko}, \citenamefont {Yurovsky},\ and\ \citenamefont
  {Olshanii}}]{rigol07}%
  \BibitemOpen
  \bibfield  {author} {\bibinfo {author} {\bibfnamefont {M.}~\bibnamefont
  {Rigol}}, \bibinfo {author} {\bibfnamefont {V.}~\bibnamefont {Dunjko}},
  \bibinfo {author} {\bibfnamefont {V.}~\bibnamefont {Yurovsky}}, \ and\
  \bibinfo {author} {\bibfnamefont {M.}~\bibnamefont {Olshanii}},\ }\href@noop
  {} {\bibfield  {journal} {\bibinfo  {journal} {Phys. Rev. Lett.}\ }\textbf
  {\bibinfo {volume} {98}},\ \bibinfo {pages} {050405} (\bibinfo {year}
  {2007})}\BibitemShut {NoStop}%
\bibitem [{\citenamefont {Iucci}\ and\ \citenamefont
  {Cazalilla}(2009)}]{iucci09}%
  \BibitemOpen
  \bibfield  {author} {\bibinfo {author} {\bibfnamefont {A.}~\bibnamefont
  {Iucci}}\ and\ \bibinfo {author} {\bibfnamefont {M.~A.}\ \bibnamefont
  {Cazalilla}},\ }\href {\doibase 10.1103/PhysRevA.80.063619} {\bibfield
  {journal} {\bibinfo  {journal} {Phys. Rev. A}\ }\textbf {\bibinfo {volume}
  {80}},\ \bibinfo {pages} {063619} (\bibinfo {year} {2009})}\BibitemShut
  {NoStop}%
\bibitem [{\citenamefont {Calabrese}\ \emph
  {et~al.}(2012{\natexlab{a}})\citenamefont {Calabrese}, \citenamefont
  {Essler},\ and\ \citenamefont {Fagotti}}]{calabrese12}%
  \BibitemOpen
  \bibfield  {author} {\bibinfo {author} {\bibfnamefont {P.}~\bibnamefont
  {Calabrese}}, \bibinfo {author} {\bibfnamefont {F.~H.~L.}\ \bibnamefont
  {Essler}}, \ and\ \bibinfo {author} {\bibfnamefont {M.}~\bibnamefont
  {Fagotti}},\ }\href {\doibase 10.1088/1742-5468/2012/07/P07016} {\bibfield
  {journal} {\bibinfo  {journal} {J. Stat. Mech.}\ ,\ \bibinfo {pages}
  {P07016}} (\bibinfo {year} {2012}{\natexlab{a}})}\BibitemShut {NoStop}%
\bibitem [{\citenamefont {Barthel}\ and\ \citenamefont
  {Schollw\"ock}(2008)}]{barthel08}%
  \BibitemOpen
  \bibfield  {author} {\bibinfo {author} {\bibfnamefont {T.}~\bibnamefont
  {Barthel}}\ and\ \bibinfo {author} {\bibfnamefont {U.}~\bibnamefont
  {Schollw\"ock}},\ }\href {\doibase 10.1103/PhysRevLett.100.100601} {\bibfield
   {journal} {\bibinfo  {journal} {Phys. Rev. Lett.}\ }\textbf {\bibinfo
  {volume} {100}},\ \bibinfo {pages} {100601} (\bibinfo {year}
  {2008})}\BibitemShut {NoStop}%
\bibitem [{\citenamefont {Cramer}\ \emph {et~al.}(2008)\citenamefont {Cramer},
  \citenamefont {Dawson}, \citenamefont {Eisert},\ and\ \citenamefont
  {Osborne}}]{cramer08}%
  \BibitemOpen
  \bibfield  {author} {\bibinfo {author} {\bibfnamefont {M.}~\bibnamefont
  {Cramer}}, \bibinfo {author} {\bibfnamefont {C.~M.}\ \bibnamefont {Dawson}},
  \bibinfo {author} {\bibfnamefont {J.}~\bibnamefont {Eisert}}, \ and\ \bibinfo
  {author} {\bibfnamefont {T.~J.}\ \bibnamefont {Osborne}},\ }\href {\doibase
  10.1103/PhysRevLett.100.030602} {\bibfield  {journal} {\bibinfo  {journal}
  {Phys. Rev. Lett.}\ }\textbf {\bibinfo {volume} {100}},\ \bibinfo {pages}
  {030602} (\bibinfo {year} {2008})}\BibitemShut {NoStop}%
\bibitem [{\citenamefont {Cramer}\ and\ \citenamefont
  {Eisert}(2010)}]{cramer10}%
  \BibitemOpen
  \bibfield  {author} {\bibinfo {author} {\bibfnamefont {M.}~\bibnamefont
  {Cramer}}\ and\ \bibinfo {author} {\bibfnamefont {J.}~\bibnamefont
  {Eisert}},\ }\href {\doibase 10.1088/1367-2630/12/5/055020} {\bibfield
  {journal} {\bibinfo  {journal} {New J. Phys.}\ }\textbf {\bibinfo {volume}
  {12}},\ \bibinfo {pages} {055020} (\bibinfo {year} {2010})}\BibitemShut
  {NoStop}%
\bibitem [{\citenamefont {Fioretto}\ and\ \citenamefont
  {Mussardo}(2010)}]{fioretto10}%
  \BibitemOpen
  \bibfield  {author} {\bibinfo {author} {\bibfnamefont {D.}~\bibnamefont
  {Fioretto}}\ and\ \bibinfo {author} {\bibfnamefont {G.}~\bibnamefont
  {Mussardo}},\ }\href {\doibase 10.1088/1367-2630/12/5/055015} {\bibfield
  {journal} {\bibinfo  {journal} {New J. Phys.}\ }\textbf {\bibinfo {volume}
  {12}},\ \bibinfo {pages} {055015} (\bibinfo {year} {2010})}\BibitemShut
  {NoStop}%
\bibitem [{\citenamefont {Calabrese}\ \emph {et~al.}(2011)\citenamefont
  {Calabrese}, \citenamefont {Essler},\ and\ \citenamefont
  {Fagotti}}]{calabrese11}%
  \BibitemOpen
  \bibfield  {author} {\bibinfo {author} {\bibfnamefont {P.}~\bibnamefont
  {Calabrese}}, \bibinfo {author} {\bibfnamefont {F.~H.~L.}\ \bibnamefont
  {Essler}}, \ and\ \bibinfo {author} {\bibfnamefont {M.}~\bibnamefont
  {Fagotti}},\ }\href {\doibase 10.1103/PhysRevLett.106.227203} {\bibfield
  {journal} {\bibinfo  {journal} {Phys. Rev. Lett.}\ }\textbf {\bibinfo
  {volume} {106}},\ \bibinfo {pages} {227203} (\bibinfo {year}
  {2011})}\BibitemShut {NoStop}%
\bibitem [{\citenamefont {Calabrese}\ \emph
  {et~al.}(2012{\natexlab{b}})\citenamefont {Calabrese}, \citenamefont
  {Essler},\ and\ \citenamefont {Fagotti}}]{calabrese12a}%
  \BibitemOpen
  \bibfield  {author} {\bibinfo {author} {\bibfnamefont {P.}~\bibnamefont
  {Calabrese}}, \bibinfo {author} {\bibfnamefont {F.~H.~L.}\ \bibnamefont
  {Essler}}, \ and\ \bibinfo {author} {\bibfnamefont {M.}~\bibnamefont
  {Fagotti}},\ }\href {\doibase 10.1088/1742-5468/2012/07/P07022} {\bibfield
  {journal} {\bibinfo  {journal} {J. Stat. Mech.}\ ,\ \bibinfo {pages}
  {P07022}} (\bibinfo {year} {2012}{\natexlab{b}})}\BibitemShut {NoStop}%
\bibitem [{\citenamefont {Caux}\ and\ \citenamefont {Konik}(2012)}]{caux12}%
  \BibitemOpen
  \bibfield  {author} {\bibinfo {author} {\bibfnamefont {J.-S.}\ \bibnamefont
  {Caux}}\ and\ \bibinfo {author} {\bibfnamefont {R.~M.}\ \bibnamefont
  {Konik}},\ }\href {\doibase 10.1103/PhysRevLett.109.175301} {\bibfield
  {journal} {\bibinfo  {journal} {Phys. Rev. Lett.}\ }\textbf {\bibinfo
  {volume} {109}},\ \bibinfo {pages} {175301} (\bibinfo {year}
  {2012})}\BibitemShut {NoStop}%
\bibitem [{\citenamefont {Collura}\ \emph {et~al.}(2013)\citenamefont
  {Collura}, \citenamefont {Sotiriadis},\ and\ \citenamefont
  {Calabrese}}]{collura13}%
  \BibitemOpen
  \bibfield  {author} {\bibinfo {author} {\bibfnamefont {M.}~\bibnamefont
  {Collura}}, \bibinfo {author} {\bibfnamefont {S.}~\bibnamefont {Sotiriadis}},
  \ and\ \bibinfo {author} {\bibfnamefont {P.}~\bibnamefont {Calabrese}},\
  }\href {\doibase 10.1103/PhysRevLett.110.245301} {\bibfield  {journal}
  {\bibinfo  {journal} {Phys. Rev. Lett.}\ }\textbf {\bibinfo {volume} {110}},\
  \bibinfo {pages} {245301} (\bibinfo {year} {2013})}\BibitemShut {NoStop}%
\bibitem [{\citenamefont {Lieb}\ and\ \citenamefont {Robinson}(1972)}]{lieb72}%
  \BibitemOpen
  \bibfield  {author} {\bibinfo {author} {\bibfnamefont {E.~H.}\ \bibnamefont
  {Lieb}}\ and\ \bibinfo {author} {\bibfnamefont {D.~W.}\ \bibnamefont
  {Robinson}},\ }\href@noop {} {\bibfield  {journal} {\bibinfo  {journal}
  {Commun. Math. Phys.}\ }\textbf {\bibinfo {volume} {28}},\ \bibinfo {pages}
  {251} (\bibinfo {year} {1972})}\BibitemShut {NoStop}%
\bibitem [{\citenamefont {Sims}\ and\ \citenamefont
  {Nachtergaele}(2010)}]{sims10}%
  \BibitemOpen
  \bibfield  {author} {\bibinfo {author} {\bibfnamefont {R.}~\bibnamefont
  {Sims}}\ and\ \bibinfo {author} {\bibfnamefont {B.}~\bibnamefont
  {Nachtergaele}},\ }\href {\doibase 10.1090/conm/529} {\emph {\bibinfo {title}
  {{Lieb-Robinson bounds in quantum many-body physics}}}},\ edited by\ \bibinfo
  {editor} {\bibfnamefont {R.}~\bibnamefont {Sims}}\ and\ \bibinfo {editor}
  {\bibfnamefont {D.}~\bibnamefont {Ueltschi}},\ \bibinfo {series} {{Entropy
  and the Quantum}}, Vol.\ \bibinfo {volume} {529}\ (\bibinfo  {publisher}
  {American Mathematical Society},\ \bibinfo {year} {2010})\BibitemShut
  {NoStop}%
\bibitem [{\citenamefont {J\"unemann}\ \emph {et~al.}(2013)\citenamefont
  {J\"unemann}, \citenamefont {Cadarso}, \citenamefont {Perez-Garcia},
  \citenamefont {Bermudez},\ and\ \citenamefont {Garcia-Ripoll}}]{juneman13}%
  \BibitemOpen
  \bibfield  {author} {\bibinfo {author} {\bibfnamefont {J.}~\bibnamefont
  {J\"unemann}}, \bibinfo {author} {\bibfnamefont {A.}~\bibnamefont {Cadarso}},
  \bibinfo {author} {\bibfnamefont {D.}~\bibnamefont {Perez-Garcia}}, \bibinfo
  {author} {\bibfnamefont {A.}~\bibnamefont {Bermudez}}, \ and\ \bibinfo
  {author} {\bibfnamefont {J.~J.}\ \bibnamefont {Garcia-Ripoll}},\ }\href
  {\doibase 10.1103/PhysRevLett.111.230404} {\bibfield  {journal} {\bibinfo
  {journal} {Phys. Rev. Lett.}\ }\textbf {\bibinfo {volume} {111}},\ \bibinfo
  {pages} {230404} (\bibinfo {year} {2013})}\BibitemShut {NoStop}%
\bibitem [{\citenamefont {Lieb}\ and\ \citenamefont
  {Vershynina}(2013)}]{lieb13}%
  \BibitemOpen
  \bibfield  {author} {\bibinfo {author} {\bibfnamefont {E.~H.}\ \bibnamefont
  {Lieb}}\ and\ \bibinfo {author} {\bibfnamefont {A.}~\bibnamefont
  {Vershynina}},\ }\href@noop {} {\bibfield  {journal} {\bibinfo  {journal}
  {arXiv:1306.0546v1}\ } (\bibinfo {year} {2013})}\BibitemShut {NoStop}%
\bibitem [{\citenamefont {Kliesch}\ \emph {et~al.}(2013)\citenamefont
  {Kliesch}, \citenamefont {Gogolin},\ and\ \citenamefont
  {Eisert}}]{kliesch13}%
  \BibitemOpen
  \bibfield  {author} {\bibinfo {author} {\bibfnamefont {M.}~\bibnamefont
  {Kliesch}}, \bibinfo {author} {\bibfnamefont {C.}~\bibnamefont {Gogolin}}, \
  and\ \bibinfo {author} {\bibfnamefont {J.}~\bibnamefont {Eisert}},\
  }\href@noop {} {\emph {\bibinfo {title} {{Lieb-Robinson bounds and the
  simulation of time evolution of local observables in lattice systems}}}},\
  edited by\ \bibinfo {editor} {\bibfnamefont {L.~D.}\ \bibnamefont {Site}}\
  and\ \bibinfo {editor} {\bibnamefont {Bach}},\ {Many-Electron Approaches in
  Physics, Chemistry and Mathematics: A Multidisciplinary View}\ (\bibinfo
  {publisher} {Springer},\ \bibinfo {year} {2013})\BibitemShut {NoStop}%
\bibitem [{\citenamefont {Poulin}(2010)}]{poulin10}%
  \BibitemOpen
  \bibfield  {author} {\bibinfo {author} {\bibfnamefont {D.}~\bibnamefont
  {Poulin}},\ }\href {\doibase 10.1103/PhysRevLett.104.190401} {\bibfield
  {journal} {\bibinfo  {journal} {Phys. Rev. Lett.}\ }\textbf {\bibinfo
  {volume} {104}},\ \bibinfo {pages} {190401} (\bibinfo {year}
  {2010})}\BibitemShut {NoStop}%
\bibitem [{\citenamefont {Bravyi}\ \emph {et~al.}(2006)\citenamefont {Bravyi},
  \citenamefont {Hastings},\ and\ \citenamefont {Verstraete}}]{bravyi06}%
  \BibitemOpen
  \bibfield  {author} {\bibinfo {author} {\bibfnamefont {S.}~\bibnamefont
  {Bravyi}}, \bibinfo {author} {\bibfnamefont {M.~B.}\ \bibnamefont
  {Hastings}}, \ and\ \bibinfo {author} {\bibfnamefont {F.}~\bibnamefont
  {Verstraete}},\ }\href {\doibase 10.1103/PhysRevLett.97.050401} {\bibfield
  {journal} {\bibinfo  {journal} {Phys. Rev. Lett.}\ }\textbf {\bibinfo
  {volume} {97}},\ \bibinfo {pages} {050401} (\bibinfo {year}
  {2006})}\BibitemShut {NoStop}%
\bibitem [{\citenamefont {Calabrese}\ and\ \citenamefont
  {Cardy}(2006)}]{calabrese06}%
  \BibitemOpen
  \bibfield  {author} {\bibinfo {author} {\bibfnamefont {P.}~\bibnamefont
  {Calabrese}}\ and\ \bibinfo {author} {\bibfnamefont {J.}~\bibnamefont
  {Cardy}},\ }\href {\doibase 10.1103/PhysRevLett.96.136801} {\bibfield
  {journal} {\bibinfo  {journal} {Phys. Rev. Lett.}\ }\textbf {\bibinfo
  {volume} {96}},\ \bibinfo {pages} {136801} (\bibinfo {year}
  {2006})}\BibitemShut {NoStop}%
\bibitem [{\citenamefont {Calabrese}\ and\ \citenamefont
  {Cardy}(2005)}]{calabrese05}%
  \BibitemOpen
  \bibfield  {author} {\bibinfo {author} {\bibfnamefont {P.}~\bibnamefont
  {Calabrese}}\ and\ \bibinfo {author} {\bibfnamefont {J.}~\bibnamefont
  {Cardy}},\ }\href {\doibase doi:10.1088/1742-5468/2005/04/P04010} {\bibfield
  {journal} {\bibinfo  {journal} {J. Stat. Mech.}\ ,\ \bibinfo {pages}
  {P04010}} (\bibinfo {year} {2005})}\BibitemShut {NoStop}%
\bibitem [{\citenamefont {Calabrese}\ and\ \citenamefont
  {Cardy}(2007)}]{calabrese07}%
  \BibitemOpen
  \bibfield  {author} {\bibinfo {author} {\bibfnamefont {P.}~\bibnamefont
  {Calabrese}}\ and\ \bibinfo {author} {\bibfnamefont {J.}~\bibnamefont
  {Cardy}},\ }\href {\doibase 10.1088/1742-5468/2007/06/P06008} {\bibfield
  {journal} {\bibinfo  {journal} {J. Stat. Mech.: T}\ ,\ \bibinfo {pages}
  {P06008}} (\bibinfo {year} {2007})}\BibitemShut {NoStop}%
\bibitem [{\citenamefont {De~Chiara}\ \emph {et~al.}(2006)\citenamefont
  {De~Chiara}, \citenamefont {Montangero}, \citenamefont {Calabrese},\ and\
  \citenamefont {Fazio}}]{dechiara06}%
  \BibitemOpen
  \bibfield  {author} {\bibinfo {author} {\bibfnamefont {G.}~\bibnamefont
  {De~Chiara}}, \bibinfo {author} {\bibfnamefont {S.}~\bibnamefont
  {Montangero}}, \bibinfo {author} {\bibfnamefont {P.}~\bibnamefont
  {Calabrese}}, \ and\ \bibinfo {author} {\bibfnamefont {R.}~\bibnamefont
  {Fazio}},\ }\href {\doibase 10.1088/1742-5468/2006/03/P03001} {\bibfield
  {journal} {\bibinfo  {journal} {J. Stat. Mech.}\ ,\ \bibinfo {pages}
  {P03001}} (\bibinfo {year} {2006})}\BibitemShut {NoStop}%
\bibitem [{\citenamefont {L{\"a}uchli}\ and\ \citenamefont
  {Kollath}(2008)}]{laeuchli08}%
  \BibitemOpen
  \bibfield  {author} {\bibinfo {author} {\bibfnamefont {A.}~\bibnamefont
  {L{\"a}uchli}}\ and\ \bibinfo {author} {\bibfnamefont {C.}~\bibnamefont
  {Kollath}},\ }\href {\doibase doi:10.1088/1742-5468/2008/05/P05018}
  {\bibfield  {journal} {\bibinfo  {journal} {J. Stat. Mech.}\ ,\ \bibinfo
  {pages} {P05018}} (\bibinfo {year} {2008})}\BibitemShut {NoStop}%
\bibitem [{\citenamefont {Manmana}\ \emph {et~al.}(2009)\citenamefont
  {Manmana}, \citenamefont {Wessel}, \citenamefont {Noack},\ and\ \citenamefont
  {Muramatsu}}]{manmana09}%
  \BibitemOpen
  \bibfield  {author} {\bibinfo {author} {\bibfnamefont {S.~R.}\ \bibnamefont
  {Manmana}}, \bibinfo {author} {\bibfnamefont {S.}~\bibnamefont {Wessel}},
  \bibinfo {author} {\bibfnamefont {R.~M.}\ \bibnamefont {Noack}}, \ and\
  \bibinfo {author} {\bibfnamefont {A.}~\bibnamefont {Muramatsu}},\ }\href
  {\doibase 10.1103/PhysRevB.79.155104} {\bibfield  {journal} {\bibinfo
  {journal} {Phys. Rev. B}\ }\textbf {\bibinfo {volume} {79}},\ \bibinfo
  {pages} {155104} (\bibinfo {year} {2009})}\BibitemShut {NoStop}%
\bibitem [{\citenamefont {Hauke}\ and\ \citenamefont
  {Tagliacozzo}(2013)}]{hauke13}%
  \BibitemOpen
  \bibfield  {author} {\bibinfo {author} {\bibfnamefont {P.}~\bibnamefont
  {Hauke}}\ and\ \bibinfo {author} {\bibfnamefont {L.}~\bibnamefont
  {Tagliacozzo}},\ }\href {\doibase 10.1103/PhysRevLett.111.207202} {\bibfield
  {journal} {\bibinfo  {journal} {Phys. Rev. Lett.}\ }\textbf {\bibinfo
  {volume} {111}},\ \bibinfo {pages} {207202} (\bibinfo {year}
  {2013})}\BibitemShut {NoStop}%
\bibitem [{\citenamefont {Eisert}\ \emph {et~al.}(2013)\citenamefont {Eisert},
  \citenamefont {van~den Worm}, \citenamefont {Manmana},\ and\ \citenamefont
  {Kastner}}]{eisert13}%
  \BibitemOpen
  \bibfield  {author} {\bibinfo {author} {\bibfnamefont {J.}~\bibnamefont
  {Eisert}}, \bibinfo {author} {\bibfnamefont {M.}~\bibnamefont {van~den
  Worm}}, \bibinfo {author} {\bibfnamefont {S.~R.}\ \bibnamefont {Manmana}}, \
  and\ \bibinfo {author} {\bibfnamefont {M.}~\bibnamefont {Kastner}},\ }\href
  {\doibase 10.1103/PhysRevLett.111.260401} {\bibfield  {journal} {\bibinfo
  {journal} {Phys. Rev. Lett.}\ }\textbf {\bibinfo {volume} {111}},\ \bibinfo
  {pages} {260401} (\bibinfo {year} {2013})}\BibitemShut {NoStop}%
\bibitem [{\citenamefont {Carleo}\ \emph {et~al.}(2014)\citenamefont {Carleo},
  \citenamefont {Becca}, \citenamefont {Sanchez-Palencia}, \citenamefont
  {Sorella},\ and\ \citenamefont {Fabrizio}}]{carleo14}%
  \BibitemOpen
  \bibfield  {author} {\bibinfo {author} {\bibfnamefont {G.}~\bibnamefont
  {Carleo}}, \bibinfo {author} {\bibfnamefont {F.}~\bibnamefont {Becca}},
  \bibinfo {author} {\bibfnamefont {L.}~\bibnamefont {Sanchez-Palencia}},
  \bibinfo {author} {\bibfnamefont {S.}~\bibnamefont {Sorella}}, \ and\
  \bibinfo {author} {\bibfnamefont {M.}~\bibnamefont {Fabrizio}},\ }\href
  {\doibase 10.1103/PhysRevA.89.031602} {\bibfield  {journal} {\bibinfo
  {journal} {Phys. Rev. A}\ }\textbf {\bibinfo {volume} {89}},\ \bibinfo
  {pages} {031602} (\bibinfo {year} {2014})}\BibitemShut {NoStop}%
\bibitem [{\citenamefont {Cheneau}\ \emph {et~al.}(2012)\citenamefont
  {Cheneau}, \citenamefont {Barmettler}, \citenamefont {Poletti}, \citenamefont
  {Endres}, \citenamefont {Schau{\ss}}, \citenamefont {Fukuhura}, \citenamefont
  {Gross}, \citenamefont {Bloch}, \citenamefont {Kollath},\ and\ \citenamefont
  {Kuhr}}]{cheneau12}%
  \BibitemOpen
  \bibfield  {author} {\bibinfo {author} {\bibfnamefont {M.}~\bibnamefont
  {Cheneau}}, \bibinfo {author} {\bibfnamefont {P.}~\bibnamefont {Barmettler}},
  \bibinfo {author} {\bibfnamefont {D.}~\bibnamefont {Poletti}}, \bibinfo
  {author} {\bibfnamefont {H.}~\bibnamefont {Endres}}, \bibinfo {author}
  {\bibfnamefont {P.}~\bibnamefont {Schau{\ss}}}, \bibinfo {author}
  {\bibfnamefont {T.}~\bibnamefont {Fukuhura}}, \bibinfo {author}
  {\bibfnamefont {C.}~\bibnamefont {Gross}}, \bibinfo {author} {\bibfnamefont
  {I.}~\bibnamefont {Bloch}}, \bibinfo {author} {\bibfnamefont
  {C.}~\bibnamefont {Kollath}}, \ and\ \bibinfo {author} {\bibfnamefont
  {S.}~\bibnamefont {Kuhr}},\ }\href {\doibase doi:10.1038/nature10748}
  {\bibfield  {journal} {\bibinfo  {journal} {Nature}\ }\textbf {\bibinfo
  {volume} {481}},\ \bibinfo {pages} {484} (\bibinfo {year}
  {2012})}\BibitemShut {NoStop}%
\bibitem [{\citenamefont {Langen}\ \emph {et~al.}(2013)\citenamefont {Langen},
  \citenamefont {Geiger}, \citenamefont {Kuhnert}, \citenamefont {Rauer},\ and\
  \citenamefont {Schmiedmayer}}]{langen13}%
  \BibitemOpen
  \bibfield  {author} {\bibinfo {author} {\bibfnamefont {T.}~\bibnamefont
  {Langen}}, \bibinfo {author} {\bibfnamefont {R.}~\bibnamefont {Geiger}},
  \bibinfo {author} {\bibfnamefont {M.}~\bibnamefont {Kuhnert}}, \bibinfo
  {author} {\bibfnamefont {B.}~\bibnamefont {Rauer}}, \ and\ \bibinfo {author}
  {\bibfnamefont {J.}~\bibnamefont {Schmiedmayer}},\ }\href {\doibase
  10.1038/nphys2739} {\bibfield  {journal} {\bibinfo  {journal} {Nature
  Physics}\ }\textbf {\bibinfo {volume} {9}},\ \bibinfo {pages} {640} (\bibinfo
  {year} {2013})}\BibitemShut {NoStop}%
\bibitem [{\citenamefont {Jurcevic}\ \emph {et~al.}(2014)\citenamefont
  {Jurcevic}, \citenamefont {Lanyon}, \citenamefont {Hauke}, \citenamefont
  {Hempel}, \citenamefont {Zoller}, \citenamefont {Blatt},\ and\ \citenamefont
  {Roos}}]{jurcevic14}%
  \BibitemOpen
  \bibfield  {author} {\bibinfo {author} {\bibfnamefont {P.}~\bibnamefont
  {Jurcevic}}, \bibinfo {author} {\bibfnamefont {B.~P.}\ \bibnamefont
  {Lanyon}}, \bibinfo {author} {\bibfnamefont {P.}~\bibnamefont {Hauke}},
  \bibinfo {author} {\bibfnamefont {C.}~\bibnamefont {Hempel}}, \bibinfo
  {author} {\bibfnamefont {P.}~\bibnamefont {Zoller}}, \bibinfo {author}
  {\bibfnamefont {R.}~\bibnamefont {Blatt}}, \ and\ \bibinfo {author}
  {\bibfnamefont {C.~F.}\ \bibnamefont {Roos}},\ }\href@noop {} {\bibfield
  {journal} {\bibinfo  {journal} {arXiv:1401.5387}\ } (\bibinfo {year}
  {2014})}\BibitemShut {NoStop}%
\bibitem [{\citenamefont {Richerme}\ \emph {et~al.}(2014)\citenamefont
  {Richerme}, \citenamefont {Gong}, \citenamefont {Lee}, \citenamefont {Senko},
  \citenamefont {Smith}, \citenamefont {Moss-Feig}, \citenamefont {Michalakis},
  \citenamefont {Gorshkov},\ and\ \citenamefont {Monroe}}]{richerme14}%
  \BibitemOpen
  \bibfield  {author} {\bibinfo {author} {\bibfnamefont {P.}~\bibnamefont
  {Richerme}}, \bibinfo {author} {\bibfnamefont {Z.-X.}\ \bibnamefont {Gong}},
  \bibinfo {author} {\bibfnamefont {A.}~\bibnamefont {Lee}}, \bibinfo {author}
  {\bibfnamefont {C.}~\bibnamefont {Senko}}, \bibinfo {author} {\bibfnamefont
  {J.}~\bibnamefont {Smith}}, \bibinfo {author} {\bibfnamefont
  {M.}~\bibnamefont {Moss-Feig}}, \bibinfo {author} {\bibfnamefont
  {S.}~\bibnamefont {Michalakis}}, \bibinfo {author} {\bibfnamefont {A.~V.}\
  \bibnamefont {Gorshkov}}, \ and\ \bibinfo {author} {\bibfnamefont
  {C.}~\bibnamefont {Monroe}},\ }\href@noop {} {\bibfield  {journal} {\bibinfo
  {journal} {arXiv:1401.5088}\ } (\bibinfo {year} {2014})}\BibitemShut
  {NoStop}%
\bibitem [{\citenamefont {Karrasch}\ \emph {et~al.}(2014)\citenamefont
  {Karrasch}, \citenamefont {Moore},\ and\ \citenamefont
  {Heidrich-Meisner}}]{karrasch14}%
  \BibitemOpen
  \bibfield  {author} {\bibinfo {author} {\bibfnamefont {C.}~\bibnamefont
  {Karrasch}}, \bibinfo {author} {\bibfnamefont {J.~E.}\ \bibnamefont {Moore}},
  \ and\ \bibinfo {author} {\bibfnamefont {F.}~\bibnamefont
  {Heidrich-Meisner}},\ }\href {\doibase 10.1103/PhysRevB.89.075139} {\bibfield
   {journal} {\bibinfo  {journal} {Phys. Rev. B}\ }\textbf {\bibinfo {volume}
  {89}},\ \bibinfo {pages} {075139} (\bibinfo {year} {2014})}\BibitemShut
  {NoStop}%
\bibitem [{\citenamefont {White}(2009)}]{white09}%
  \BibitemOpen
  \bibfield  {author} {\bibinfo {author} {\bibfnamefont {S.~R.}\ \bibnamefont
  {White}},\ }\href {\doibase 10.1103/PhysRevLett.102.190601} {\bibfield
  {journal} {\bibinfo  {journal} {Phys. Rev. Lett.}\ }\textbf {\bibinfo
  {volume} {102}},\ \bibinfo {pages} {190601} (\bibinfo {year}
  {2009})}\BibitemShut {NoStop}%
\bibitem [{\citenamefont {Stoudenmire}\ and\ \citenamefont
  {White}(2010)}]{stoudenmire10}%
  \BibitemOpen
  \bibfield  {author} {\bibinfo {author} {\bibfnamefont {E.~M.}\ \bibnamefont
  {Stoudenmire}}\ and\ \bibinfo {author} {\bibfnamefont {S.~R.}\ \bibnamefont
  {White}},\ }\href {\doibase doi:10.1088/1367-2630/12/5/055026} {\bibfield
  {journal} {\bibinfo  {journal} {New J. Phys.}\ }\textbf {\bibinfo {volume}
  {12}},\ \bibinfo {pages} {055026} (\bibinfo {year} {2010})}\BibitemShut
  {NoStop}%
\bibitem [{Note1()}]{Note1}%
  \BibitemOpen
  \bibinfo {note} {Note that the system is \protect \textit {not} connected to
  a heat bath during the time evolution and energy $\protect \mathrm {Tr}[H
  \rho (t)]$ is conserved.}\BibitemShut {Stop}%
\bibitem [{\citenamefont {Barmettler}\ \emph {et~al.}(2012)\citenamefont
  {Barmettler}, \citenamefont {Poletti}, \citenamefont {Cheneau},\ and\
  \citenamefont {Kollath}}]{barmettler12}%
  \BibitemOpen
  \bibfield  {author} {\bibinfo {author} {\bibfnamefont {P.}~\bibnamefont
  {Barmettler}}, \bibinfo {author} {\bibfnamefont {D.}~\bibnamefont {Poletti}},
  \bibinfo {author} {\bibfnamefont {M.}~\bibnamefont {Cheneau}}, \ and\
  \bibinfo {author} {\bibfnamefont {C.}~\bibnamefont {Kollath}},\ }\href
  {\doibase 10.1103/PhysRevA.85.053625} {\bibfield  {journal} {\bibinfo
  {journal} {Phys. Rev. A}\ }\textbf {\bibinfo {volume} {85}},\ \bibinfo
  {pages} {053625} (\bibinfo {year} {2012})}\BibitemShut {NoStop}%
\bibitem [{\citenamefont {Caux}\ and\ \citenamefont {Essler}(2013)}]{caux13}%
  \BibitemOpen
  \bibfield  {author} {\bibinfo {author} {\bibfnamefont {J.-S.}\ \bibnamefont
  {Caux}}\ and\ \bibinfo {author} {\bibfnamefont {F.~H.~L.}\ \bibnamefont
  {Essler}},\ }\href {\doibase 10.1103/PhysRevLett.110.257203} {\bibfield
  {journal} {\bibinfo  {journal} {Phys. Rev. Lett.}\ }\textbf {\bibinfo
  {volume} {110}},\ \bibinfo {pages} {257203} (\bibinfo {year}
  {2013})}\BibitemShut {NoStop}%
\bibitem [{\citenamefont {Mossel}\ and\ \citenamefont {Caux}(2012)}]{mossel12}%
  \BibitemOpen
  \bibfield  {author} {\bibinfo {author} {\bibfnamefont {J.}~\bibnamefont
  {Mossel}}\ and\ \bibinfo {author} {\bibfnamefont {J.-S.}\ \bibnamefont
  {Caux}},\ }\href {\doibase 10.1088/1751-8113/45/25/255001} {\bibfield
  {journal} {\bibinfo  {journal} {J. Phys. A}\ }\textbf {\bibinfo {volume}
  {45}},\ \bibinfo {pages} {255001} (\bibinfo {year} {2012})}\BibitemShut
  {NoStop}%
\bibitem [{\citenamefont {Demler}\ and\ \citenamefont
  {Tsvelik}(2012)}]{demler12}%
  \BibitemOpen
  \bibfield  {author} {\bibinfo {author} {\bibfnamefont {E.}~\bibnamefont
  {Demler}}\ and\ \bibinfo {author} {\bibfnamefont {A.~M.}\ \bibnamefont
  {Tsvelik}},\ }\href {\doibase 10.1103/PhysRevB.86.115448} {\bibfield
  {journal} {\bibinfo  {journal} {Phys. Rev. B}\ }\textbf {\bibinfo {volume}
  {86}},\ \bibinfo {pages} {115448} (\bibinfo {year} {2012})}\BibitemShut
  {NoStop}%
\bibitem [{\citenamefont {Grabowski}\ and\ \citenamefont
  {Mathieu}(1995)}]{grabowski95}%
  \BibitemOpen
  \bibfield  {author} {\bibinfo {author} {\bibfnamefont {M.~P.}\ \bibnamefont
  {Grabowski}}\ and\ \bibinfo {author} {\bibfnamefont {P.}~\bibnamefont
  {Mathieu}},\ }\href {\doibase 10.1006/aphy.1995.1101} {\bibfield  {journal}
  {\bibinfo  {journal} {Ann. Phys.}\ }\textbf {\bibinfo {volume} {243}},\
  \bibinfo {pages} {299} (\bibinfo {year} {1995})}\BibitemShut {NoStop}%
\bibitem [{\citenamefont {Prosen}\ and\ \citenamefont
  {Ilievski}(2013)}]{prosen13}%
  \BibitemOpen
  \bibfield  {author} {\bibinfo {author} {\bibfnamefont {T.}~\bibnamefont
  {Prosen}}\ and\ \bibinfo {author} {\bibfnamefont {E.}~\bibnamefont
  {Ilievski}},\ }\href {\doibase 10.1103/PhysRevLett.111.057203} {\bibfield
  {journal} {\bibinfo  {journal} {Phys. Rev. Lett.}\ }\textbf {\bibinfo
  {volume} {111}},\ \bibinfo {pages} {057203} (\bibinfo {year}
  {2013})}\BibitemShut {NoStop}%
\bibitem [{\citenamefont {Pozsgay}(2013)}]{pozsgay13}%
  \BibitemOpen
  \bibfield  {author} {\bibinfo {author} {\bibfnamefont {B.}~\bibnamefont
  {Pozsgay}},\ }\href {\doibase 10.1088/1742-5468/2013/07/P07003} {\bibfield
  {journal} {\bibinfo  {journal} {J. Stat. Mech.}\ ,\ \bibinfo {pages}
  {P07003}} (\bibinfo {year} {2013})}\BibitemShut {NoStop}%
\bibitem [{\citenamefont {Fagotti}\ and\ \citenamefont
  {Essler}(2013{\natexlab{a}})}]{fagotti13}%
  \BibitemOpen
  \bibfield  {author} {\bibinfo {author} {\bibfnamefont {M.}~\bibnamefont
  {Fagotti}}\ and\ \bibinfo {author} {\bibfnamefont {F.~H.~L.}\ \bibnamefont
  {Essler}},\ }\href {\doibase 10.1088/1742-5468/2013/07/P07012} {\bibfield
  {journal} {\bibinfo  {journal} {J. Stat. Mech.}\ ,\ \bibinfo {pages}
  {P07012}} (\bibinfo {year} {2013}{\natexlab{a}})}\BibitemShut {NoStop}%
\bibitem [{\citenamefont {Fagotti}\ \emph {et~al.}(2014)\citenamefont
  {Fagotti}, \citenamefont {Collura}, \citenamefont {Essler},\ and\
  \citenamefont {Calabrese}}]{fagotti14}%
  \BibitemOpen
  \bibfield  {author} {\bibinfo {author} {\bibfnamefont {M.}~\bibnamefont
  {Fagotti}}, \bibinfo {author} {\bibfnamefont {M.}~\bibnamefont {Collura}},
  \bibinfo {author} {\bibfnamefont {F.~H.~L.}\ \bibnamefont {Essler}}, \ and\
  \bibinfo {author} {\bibfnamefont {P.}~\bibnamefont {Calabrese}},\ }\href
  {\doibase 10.1103/PhysRevB.89.125101} {\bibfield  {journal} {\bibinfo
  {journal} {Phys. Rev. B}\ }\textbf {\bibinfo {volume} {89}},\ \bibinfo
  {pages} {125101} (\bibinfo {year} {2014})}\BibitemShut {NoStop}%
\bibitem [{\citenamefont {Wouters}\ \emph {et~al.}(2014)\citenamefont
  {Wouters}, \citenamefont {Brockmann}, \citenamefont {De~Nardis},
  \citenamefont {Fioretto},\ and\ \citenamefont {Caux}}]{wouters14}%
  \BibitemOpen
  \bibfield  {author} {\bibinfo {author} {\bibfnamefont {B.}~\bibnamefont
  {Wouters}}, \bibinfo {author} {\bibfnamefont {M.}~\bibnamefont {Brockmann}},
  \bibinfo {author} {\bibfnamefont {J.}~\bibnamefont {De~Nardis}}, \bibinfo
  {author} {\bibfnamefont {D.}~\bibnamefont {Fioretto}}, \ and\ \bibinfo
  {author} {\bibfnamefont {J.-S.}\ \bibnamefont {Caux}},\ }\href@noop {}
  {\bibfield  {journal} {\bibinfo  {journal} {arXiv:1405.0172}\ } (\bibinfo
  {year} {2014})}\BibitemShut {NoStop}%
\bibitem [{\citenamefont {Poszgay}\ \emph {et~al.}(2014)\citenamefont
  {Poszgay}, \citenamefont {Mesty{\'a}n}, \citenamefont {Werner}, \citenamefont
  {Kormos}, \citenamefont {Zar{\'a}nd},\ and\ \citenamefont
  {Tak{\'a}s}}]{poszgay14}%
  \BibitemOpen
  \bibfield  {author} {\bibinfo {author} {\bibfnamefont {B.}~\bibnamefont
  {Poszgay}}, \bibinfo {author} {\bibfnamefont {M.}~\bibnamefont
  {Mesty{\'a}n}}, \bibinfo {author} {\bibfnamefont {M.~A.}\ \bibnamefont
  {Werner}}, \bibinfo {author} {\bibfnamefont {M.}~\bibnamefont {Kormos}},
  \bibinfo {author} {\bibfnamefont {G.}~\bibnamefont {Zar{\'a}nd}}, \ and\
  \bibinfo {author} {\bibfnamefont {G.}~\bibnamefont {Tak{\'a}s}},\ }\href@noop
  {} {\bibfield  {journal} {\bibinfo  {journal} {arXiv:1405.2843}\ } (\bibinfo
  {year} {2014})}\BibitemShut {NoStop}%
\bibitem [{\citenamefont {Pozsgay}(2011)}]{pozsgay11}%
  \BibitemOpen
  \bibfield  {author} {\bibinfo {author} {\bibfnamefont {B.}~\bibnamefont
  {Pozsgay}},\ }\href {\doibase 10.1088/1742-5468/2011/01/P01011} {\bibfield
  {journal} {\bibinfo  {journal} {J. Stat. Mech.}\ ,\ \bibinfo {pages}
  {P01011}} (\bibinfo {year} {2011})}\BibitemShut {NoStop}%
\bibitem [{\citenamefont {Cassidy}\ \emph {et~al.}(2011)\citenamefont
  {Cassidy}, \citenamefont {Clark},\ and\ \citenamefont {Rigol}}]{cassidy11}%
  \BibitemOpen
  \bibfield  {author} {\bibinfo {author} {\bibfnamefont {A.~C.}\ \bibnamefont
  {Cassidy}}, \bibinfo {author} {\bibfnamefont {C.~W.}\ \bibnamefont {Clark}},
  \ and\ \bibinfo {author} {\bibfnamefont {M.}~\bibnamefont {Rigol}},\ }\href
  {\doibase 10.1103/PhysRevLett.106.140405} {\bibfield  {journal} {\bibinfo
  {journal} {Phys. Rev. Lett.}\ }\textbf {\bibinfo {volume} {106}},\ \bibinfo
  {pages} {140405} (\bibinfo {year} {2011})}\BibitemShut {NoStop}%
\bibitem [{\citenamefont {Takahashi}\ and\ \citenamefont
  {Suzuki}(1972)}]{takahashi72}%
  \BibitemOpen
  \bibfield  {author} {\bibinfo {author} {\bibfnamefont {M.}~\bibnamefont
  {Takahashi}}\ and\ \bibinfo {author} {\bibfnamefont {M.}~\bibnamefont
  {Suzuki}},\ }\href {\doibase 10.1143/PTP.48.2187} {\bibfield  {journal}
  {\bibinfo  {journal} {Prog. Theor. Phys.}\ }\textbf {\bibinfo {volume}
  {48}},\ \bibinfo {pages} {2187} (\bibinfo {year} {1972})}\BibitemShut
  {NoStop}%
\bibitem [{\citenamefont {Fagotti}\ and\ \citenamefont
  {Essler}(2013{\natexlab{b}})}]{fagotti13b}%
  \BibitemOpen
  \bibfield  {author} {\bibinfo {author} {\bibfnamefont {M.}~\bibnamefont
  {Fagotti}}\ and\ \bibinfo {author} {\bibfnamefont {F.~H.~L.}\ \bibnamefont
  {Essler}},\ }\href {\doibase 10.1103/PhysRevB.87.245107} {\bibfield
  {journal} {\bibinfo  {journal} {Phys. Rev. B}\ }\textbf {\bibinfo {volume}
  {87}},\ \bibinfo {pages} {245107} (\bibinfo {year}
  {2013}{\natexlab{b}})}\BibitemShut {NoStop}%
\bibitem [{\citenamefont {Vidal}(2003)}]{vidal03}%
  \BibitemOpen
  \bibfield  {author} {\bibinfo {author} {\bibfnamefont {G.}~\bibnamefont
  {Vidal}},\ }\href {\doibase 10.1103/PhysRevLett.91.147902} {\bibfield
  {journal} {\bibinfo  {journal} {Phys. Rev. Lett.}\ }\textbf {\bibinfo
  {volume} {91}},\ \bibinfo {pages} {147902} (\bibinfo {year}
  {2003})}\BibitemShut {NoStop}%
\bibitem [{\citenamefont {Feiguin}\ and\ \citenamefont
  {White}(2005)}]{feiguin05b}%
  \BibitemOpen
  \bibfield  {author} {\bibinfo {author} {\bibfnamefont {A.~E.}\ \bibnamefont
  {Feiguin}}\ and\ \bibinfo {author} {\bibfnamefont {S.~R.}\ \bibnamefont
  {White}},\ }\href {\doibase 10.1103/PhysRevB.72.220401} {\bibfield  {journal}
  {\bibinfo  {journal} {Phys. Rev. B}\ }\textbf {\bibinfo {volume} {72}},\
  \bibinfo {pages} {220401} (\bibinfo {year} {2005})}\BibitemShut {NoStop}%
\bibitem [{\citenamefont {Barthel}\ \emph {et~al.}(2009)\citenamefont
  {Barthel}, \citenamefont {Schollw\"ock},\ and\ \citenamefont
  {White}}]{barthel09}%
  \BibitemOpen
  \bibfield  {author} {\bibinfo {author} {\bibfnamefont {T.}~\bibnamefont
  {Barthel}}, \bibinfo {author} {\bibfnamefont {U.}~\bibnamefont
  {Schollw\"ock}}, \ and\ \bibinfo {author} {\bibfnamefont {S.~R.}\
  \bibnamefont {White}},\ }\href {\doibase 10.1103/PhysRevB.79.245101}
  {\bibfield  {journal} {\bibinfo  {journal} {Phys. Rev. B}\ }\textbf {\bibinfo
  {volume} {79}},\ \bibinfo {pages} {245101} (\bibinfo {year}
  {2009})}\BibitemShut {NoStop}%
\bibitem [{\citenamefont {Karrasch}\ \emph {et~al.}(2012)\citenamefont
  {Karrasch}, \citenamefont {Bardarson},\ and\ \citenamefont
  {Moore}}]{karrasch12}%
  \BibitemOpen
  \bibfield  {author} {\bibinfo {author} {\bibfnamefont {C.}~\bibnamefont
  {Karrasch}}, \bibinfo {author} {\bibfnamefont {J.~H.}\ \bibnamefont
  {Bardarson}}, \ and\ \bibinfo {author} {\bibfnamefont {J.~E.}\ \bibnamefont
  {Moore}},\ }\href {\doibase 10.1103/PhysRevLett.108.227206} {\bibfield
  {journal} {\bibinfo  {journal} {Phys. Rev. Lett.}\ }\textbf {\bibinfo
  {volume} {108}},\ \bibinfo {pages} {227206} (\bibinfo {year}
  {2012})}\BibitemShut {NoStop}%
\bibitem [{\citenamefont {Barthel}(2013)}]{barthel13}%
  \BibitemOpen
  \bibfield  {author} {\bibinfo {author} {\bibfnamefont {T.}~\bibnamefont
  {Barthel}},\ }\href {\doibase doi:10.1088/1367-2630/15/7/073010} {\bibfield
  {journal} {\bibinfo  {journal} {New J. Phys.}\ }\textbf {\bibinfo {volume}
  {15}},\ \bibinfo {pages} {073010} (\bibinfo {year} {2013})}\BibitemShut
  {NoStop}%
\bibitem [{\citenamefont {Karrasch}\ \emph {et~al.}(2013)\citenamefont
  {Karrasch}, \citenamefont {Bardarson},\ and\ \citenamefont
  {Moore}}]{karrasch13b}%
  \BibitemOpen
  \bibfield  {author} {\bibinfo {author} {\bibfnamefont {C.}~\bibnamefont
  {Karrasch}}, \bibinfo {author} {\bibfnamefont {J.~H.}\ \bibnamefont
  {Bardarson}}, \ and\ \bibinfo {author} {\bibfnamefont {J.~E.}\ \bibnamefont
  {Moore}},\ }\href {\doibase doi:10.1088/1367-2630/15/8/083031} {\bibfield
  {journal} {\bibinfo  {journal} {New J. Phys.}\ }\textbf {\bibinfo {volume}
  {15}},\ \bibinfo {pages} {083031} (\bibinfo {year} {2013})}\BibitemShut
  {NoStop}%
\bibitem [{\citenamefont {{M. Zwolak and G. Vidal}}(2004)}]{zwolak04}%
  \BibitemOpen
  \bibfield  {author} {\bibinfo {author} {\bibnamefont {{M. Zwolak and G.
  Vidal}}},\ }\href {\doibase 10.1103/PhysRevLett.93.207205} {\bibfield
  {journal} {\bibinfo  {journal} {Phys. Rev. Lett.}\ }\textbf {\bibinfo
  {volume} {93}},\ \bibinfo {pages} {207205} (\bibinfo {year}
  {2004})}\BibitemShut {NoStop}%
\bibitem [{\citenamefont {Verstraete}\ \emph {et~al.}(2004)\citenamefont
  {Verstraete}, \citenamefont {Garcia-Ripoll},\ and\ \citenamefont
  {Cirac}}]{verstraete04b}%
  \BibitemOpen
  \bibfield  {author} {\bibinfo {author} {\bibfnamefont {F.}~\bibnamefont
  {Verstraete}}, \bibinfo {author} {\bibfnamefont {J.~J.}\ \bibnamefont
  {Garcia-Ripoll}}, \ and\ \bibinfo {author} {\bibfnamefont {J.~I.}\
  \bibnamefont {Cirac}},\ }\href {\doibase 10.1103/PhysRevLett.93.207204}
  {\bibfield  {journal} {\bibinfo  {journal} {Phys. Rev. Lett.}\ }\textbf
  {\bibinfo {volume} {93}},\ \bibinfo {pages} {207204} (\bibinfo {year}
  {2004})}\BibitemShut {NoStop}%
\bibitem [{\citenamefont {Prosen}\ and\ \citenamefont {\ifmmode \check{Z}\else
  \v{Z}\fi{}nidari\ifmmode~\check{c}\else \v{c}\fi{}}(2010)}]{prosen10}%
  \BibitemOpen
  \bibfield  {author} {\bibinfo {author} {\bibfnamefont {T.}~\bibnamefont
  {Prosen}}\ and\ \bibinfo {author} {\bibfnamefont {M.}~\bibnamefont {\ifmmode
  \check{Z}\else \v{Z}\fi{}nidari\ifmmode~\check{c}\else \v{c}\fi{}}},\ }\href
  {\doibase 10.1088/1742-5468/2009/02/P02035} {\bibfield  {journal} {\bibinfo
  {journal} {J. Stat. Mech.}\ ,\ \bibinfo {pages} {P07020}} (\bibinfo {year}
  {2010})}\BibitemShut {NoStop}%
\bibitem [{\citenamefont {Bursill}\ \emph {et~al.}(1996)\citenamefont
  {Bursill}, \citenamefont {Xiang},\ and\ \citenamefont {Gehring}}]{bursill96}%
  \BibitemOpen
  \bibfield  {author} {\bibinfo {author} {\bibfnamefont {R.~J.}\ \bibnamefont
  {Bursill}}, \bibinfo {author} {\bibfnamefont {T.}~\bibnamefont {Xiang}}, \
  and\ \bibinfo {author} {\bibfnamefont {G.~A.}\ \bibnamefont {Gehring}},\
  }\href {\doibase 10.1088/0953-8984/8/40/003} {\bibfield  {journal} {\bibinfo
  {journal} {J. Phys.: Condens. Matter}\ }\textbf {\bibinfo {volume} {8}},\
  \bibinfo {pages} {L583} (\bibinfo {year} {1996})}\BibitemShut {NoStop}%
\bibitem [{\citenamefont {Wang}\ and\ \citenamefont {Xiang}(1997)}]{wang97}%
  \BibitemOpen
  \bibfield  {author} {\bibinfo {author} {\bibfnamefont {X.}~\bibnamefont
  {Wang}}\ and\ \bibinfo {author} {\bibfnamefont {T.}~\bibnamefont {Xiang}},\
  }\href {\doibase 10.1103/PhysRevB.56.5061} {\bibfield  {journal} {\bibinfo
  {journal} {Phys. Rev. B}\ }\textbf {\bibinfo {volume} {56}},\ \bibinfo
  {pages} {5061} (\bibinfo {year} {1997})}\BibitemShut {NoStop}%
\bibitem [{\citenamefont {Sirker}\ and\ \citenamefont
  {Kl\"umper}(2005)}]{sirker05}%
  \BibitemOpen
  \bibfield  {author} {\bibinfo {author} {\bibfnamefont {J.}~\bibnamefont
  {Sirker}}\ and\ \bibinfo {author} {\bibfnamefont {A.}~\bibnamefont
  {Kl\"umper}},\ }\href {\doibase 10.1103/PhysRevB.71.241101} {\bibfield
  {journal} {\bibinfo  {journal} {Phys. Rev. B}\ }\textbf {\bibinfo {volume}
  {71}},\ \bibinfo {pages} {241101} (\bibinfo {year} {2005})}\BibitemShut
  {NoStop}%
\bibitem [{Note2()}]{Note2}%
  \BibitemOpen
  \bibinfo {note} {It is possible to evaluate observables that commute with the
  projection operator using the CPS rather than the METTS. The real time
  evolution, though, prohibits us from taking advantage of this improved
  measurement scheme.}\BibitemShut {Stop}%
\bibitem [{\citenamefont {Gobert}\ \emph {et~al.}(2005)\citenamefont {Gobert},
  \citenamefont {Kollath}, \citenamefont {Schollw\"ock},\ and\ \citenamefont
  {Sch\"utz}}]{gobert05}%
  \BibitemOpen
  \bibfield  {author} {\bibinfo {author} {\bibfnamefont {D.}~\bibnamefont
  {Gobert}}, \bibinfo {author} {\bibfnamefont {C.}~\bibnamefont {Kollath}},
  \bibinfo {author} {\bibfnamefont {U.}~\bibnamefont {Schollw\"ock}}, \ and\
  \bibinfo {author} {\bibfnamefont {G.}~\bibnamefont {Sch\"utz}},\ }\href
  {\doibase 10.1103/PhysRevE.71.036102} {\bibfield  {journal} {\bibinfo
  {journal} {Phys. Rev. E}\ }\textbf {\bibinfo {volume} {71}},\ \bibinfo
  {pages} {036102} (\bibinfo {year} {2005})}\BibitemShut {NoStop}%
\bibitem [{\citenamefont {Bonnes}\ and\ \citenamefont
  {L{\"a}uchli}(2014)}]{bonnes14}%
  \BibitemOpen
  \bibfield  {author} {\bibinfo {author} {\bibfnamefont {L.}~\bibnamefont
  {Bonnes}}\ and\ \bibinfo {author} {\bibfnamefont {A.~M.}\ \bibnamefont
  {L{\"a}uchli}},\ }\href@noop {} {\bibfield  {journal} {\bibinfo  {journal}
  {(in preparation)}\ } (\bibinfo {year} {2014})}\BibitemShut {NoStop}%
\bibitem [{\citenamefont {Ganahl}\ \emph {et~al.}(2012)\citenamefont {Ganahl},
  \citenamefont {Rabel}, \citenamefont {Essler},\ and\ \citenamefont
  {Evertz}}]{ganahl12}%
  \BibitemOpen
  \bibfield  {author} {\bibinfo {author} {\bibfnamefont {M.}~\bibnamefont
  {Ganahl}}, \bibinfo {author} {\bibfnamefont {E.}~\bibnamefont {Rabel}},
  \bibinfo {author} {\bibfnamefont {F.~H.~L.}\ \bibnamefont {Essler}}, \ and\
  \bibinfo {author} {\bibfnamefont {H.~G.}\ \bibnamefont {Evertz}},\ }\href
  {\doibase 10.1103/PhysRevLett.108.077206} {\bibfield  {journal} {\bibinfo
  {journal} {Phys. Rev. Lett.}\ }\textbf {\bibinfo {volume} {108}},\ \bibinfo
  {pages} {077206} (\bibinfo {year} {2012})}\BibitemShut {NoStop}%
\end{thebibliography}

%

\cleardoublepage

\onecolumngrid



\section*{Supplementary material}

# Supplementary Material for "Light-cone" dynamics after quantum quenches in spin chains

Lars Bonnes,[1, *] Fabian H. L. Essler,[2] and Andreas M. Läuchli[1]

[1]*Institute for Theoretical Physics, University of Innsbruck, A-6020 Innsbruck, Austria.*
[2]*The Rudolf Peierls Centre for Theoretical Physics, Oxford University, Oxford OX1 3NP, UK*

(Dated: August 29, 2014)

## I. GENERALIZED THERMODYNAMIC BETHE ANSATZ

It is by now widely accepted, that at late time after the quench local properties of the spin-1/2 Heisenberg XXZ chain are described by a generalized Gibbs ensemble[1,2]. The latter is constructed from the local integrals of motion $\{I_n\}$, where we take $I_1$ to be equal to the Hamiltonian. The set $\{I_n\}$ supposedly contains both the "ultra-local" integrals of motion obtained by taking logarithmic derivatives of the transfer matrix at the shift-point[3,4], and the "quasi-local" operators discovered recently[5]. The GGE density matrix is of the form

$$\rho_{\text{GGE}} = \frac{1}{Z_{\text{GGE}}} e^{-\sum_{n=1} \lambda_n I_n}, \quad (1)$$

where $Z_{\text{GGE}} = \text{Tr} e^{-\sum_{n=1} \lambda_n I_n}$ and the Lagrange multipliers $\lambda_n$ are fixed by the requirements

$$i_n \equiv \lim_{L\to\infty} \frac{1}{L} \text{Tr}[\rho(t=0) I_n] = \lim_{L\to\infty} \frac{1}{L} \text{Tr}[\rho_{\text{GGE}} I_n] . \quad (2)$$

A practical way of constructing the GGE density matrix proceeds by first retaining a given number $n_0$ of the most local conservation laws, giving rise to a "truncated GGE"[6] in the thermodynamic limit, and then increasing this number until the quantity of interest ceases to depend on $n_0$. In the thermodynamic limit the density matrix $\rho_{\text{GGE}}$ is dominated by a saddle point, characterized as a the minimum of the generalized thermodynamic potential

$$G = \sum_{n=1}^{\infty} \lambda_n H^{(n)} - S, \quad (3)$$

where $S$ is the entropy. For integrable models this saddle point macro-state can be determined by a generalized thermodynamic Bethe Ansatz[7]. The necessary analysis is very similar to the one for the thermal case, which is discussed in detail in the monograph [8]. In the following we first summarize some results of the analysis for the thermal case, and then present the modifications necessary to describe the GGE.

### A. Structure of eigenstates

In order to conform to the notations of Ref. [8] we consider the Hamiltonian in the form

$$H(J, \Delta) = -J \sum_{j=1}^{L} S_j^x S_{j+1}^x + S_j^y S_{j+1}^y + \Delta S_j^z S_{j+1}^z, \quad (4)$$

and parametrize the anisotropy as

$$\cos\gamma = -\Delta . \quad (5)$$

We note that $H(J, \Delta)$ and $H(-J, -\Delta)$ are unitarily equivalent. Eigenstates of (4) can be labelled by $N$ complex spectral parameters $x_j$, where $N$ is the number of down spins

$$|x_1, \ldots, x_N\rangle. \quad (6)$$

Energy and momentum of these states are[8]

$$E = \sum_{j=1}^{N} J \frac{\sin^2 \gamma}{\cos\gamma - \cosh\gamma x_j} - \frac{N\Delta J}{4} ,$$

$$P = \sum_{j=1}^{N} i \ln \left[ -\frac{\sinh\frac{\gamma}{2}(x_j + i)}{\sinh\frac{\gamma}{2}(x_j - i)} \right] \equiv \sum_{j=1}^{N} p^{(0)}(x_j) . \quad (7)$$

Imposing periodic boundary conditions leads to the Bethe Ansatz equations

$$\left[ \frac{\sinh\frac{\gamma}{2}(x_j + i)}{\sinh\frac{\gamma}{2}(x_j - i)} \right]^L = \prod_{l\neq j}^{N} \left[ \frac{\sinh\frac{\gamma}{2}(x_j - x_l + 2i)}{\sinh\frac{\gamma}{2}(x_j - x_l - 2i)} \right],$$
$$j = 1, 2, \ldots, N. \quad (8)$$

An important feature of integrable models like the Heisenberg XXZ chain is that excitations over the ground state are composed of different types of "elementary" excitations, corresponding to particular "string" patterns of rapidities in the complex plane, the precise structure of which depends on the anisotropy parameter $\Delta$[8]. In order to keep things simple, in the following we focus on particular values of $\Delta$

$$\Delta_\ell = -\cos(\pi/\ell) , \quad \ell = 1, 2, \ldots. \quad (9)$$

For these values there are $\ell$ types of excitations[8]

- positive parity 1-strings, corresponding to real roots $x_j^1$ of (8).

- negative parity 1-strings, corresponding to roots $x_j^\ell$ of (8) with imaginary part equal to $\ell$.

- positive parity $n$-strings for $2 \leq n \leq \ell - 1$. These correspond to solutions of (8) such that

$$x_{\alpha,j}^n = x_\alpha^n + i(n+1-2j), \ j = 1,\ldots,n, \ x_\alpha^n \in \mathbb{R}. \quad (10)$$

Substituting this "string hypothesis" back in the Bethe Ansatz equations (8) and taking the logarithm leads to a set of coupled equations for the string centres known as "discrete Takahashi equations" (DTE) for a solution with $M_j$ strings of type $j$

$$L\theta_j(x_\alpha^j) - \sum_{k=1}^\ell \sum_{\beta=1}^{M_k} \Theta_{j,k}(x_\alpha^j - x_\beta^k) = 2\pi I_\alpha^j,$$
$$\alpha = 1,\ldots,M_j, \quad j = 1,\ldots,\ell. \quad (11)$$

Here $I_\alpha^j$ are integer or half-odd integer numbers,

$$\theta_j(x) = \begin{cases} 2\arctan\left[\cot\left(\frac{\pi j}{2\ell}\right)\tanh\left(\frac{\pi x}{2\ell}\right)\right] & \text{if } j \neq \ell \\ -2\arctan\left[\tan\left(\frac{\pi}{2\ell}\right)\tanh\left(\frac{\pi x}{2\ell}\right)\right] & \text{if } j = \ell. \end{cases}$$

$$\Theta_{j,k}(x) = \theta_{k-j}(x) + 2\sum_{r=1}^{j-1}\theta_{k-j+2r}(x) + \theta_{k+j}(x),$$
$$\text{if } j \leq k < \ell, \quad (12)$$

and $\Theta_{j,k}(x) = \Theta_{k,j}(x)$, $\Theta_{j,\ell}(x) = -\Theta_{j,\ell-1}(x)$. Energy and momentum of solutions of the DTE are given by

$$E = -N\frac{J\Delta}{4} + \sum_{j=1}^\ell \sum_{\alpha=1}^{M_j} \epsilon_j^{(0)}(x_\alpha^j),$$
$$P = \sum_{j=1}^\ell \sum_{\alpha=1}^{M_j} \frac{2\pi I_\alpha^j}{L}, \quad (13)$$

where the bare energies are of the form

$$\epsilon_j^{(0)}(x) = -\frac{2\pi J \sin\gamma}{\gamma} a_j(x), \quad (14)$$

$$a_j(x) = \frac{1}{2\pi}\frac{\gamma \sin(\gamma q_j)}{\cosh(\gamma x) + \cos(\gamma q_j)}. \quad (15)$$

Here the parameters $q_j$ are given by

$$q_j = \ell - j, \ j \leq \ell - 1, \ q_\ell = -1. \quad (16)$$

### B. TBA equations for the XXZ chain at finite temperature

At temperature $T > 0$ the state of thermodynamic equilibirum is characterized by root densities $\rho_j^{p,h}$ describing the distributions of "particles" and "holes" of these different types of elementary excitations. The densities are determined from the following systems of equations[8]

$$\frac{\epsilon_j(x)}{T} = \frac{\epsilon_j^{(0)}(x)}{T} + \sum_{k=1}^\ell \text{sgn}(q_k)T_{jk} * \ln\left[1 + e^{-\epsilon_k/T}\right]\bigg|_x,$$

$$a_j(x) = \text{sgn}(q_j)[\rho_j^p(x) + \rho_j^h(x)] + \sum_k T_{jk} * \rho_k^p\bigg|_x, \quad (17)$$

where the dressed energies $\epsilon_j(x)$ are defined as

$$\epsilon_j(x) = T \ln\left[\frac{\rho_j^h(x)}{\rho_j^p(x)}\right]. \quad (18)$$

The operation $*$ denotes convolution, e.g.

$$T_{jk} * \rho_k^p\bigg|_x = \int_{-\infty}^\infty dy \ T_{jk}(x-y) \ \rho_k^p(y), \quad (19)$$

and the Fourier transforms of the integral kernels are given by ($T_{jk}(x) = T_{kj}(x)$)

$$\tilde{T}_{jk}(\omega) = \delta_{j,\ell-1}\delta_{k,\ell} - \delta_{j,k} + 2\text{sgn}(q_j)\coth(\omega)$$
$$\times \frac{\sinh((\ell-|q_j|)\omega)\sinh(q_k\omega)}{\sinh(\ell\omega)}, \ j \leq k. \quad (20)$$

Given a solution of the TBA equations (17) we can for example calculate the free energy per site

$$f = -\frac{J\Delta}{4} - \sum_n \int dx \ a_n(x) \ln\left[1 + e^{-\epsilon_n(x)/T}\right] \text{sgn}(q_n). \quad (21)$$

### C. Generalized Gibbs Ensembles

The analysis for the GGE closely parallels the thermal case. The Bethe Ansatz states are simultaneous eigenstates of the Hamiltonian and the higher conservation laws, and as a result we have ($m = 2, 3, \ldots$)

$$I_m |x_1,\ldots,x_N\rangle = \left[\sum_{j=1}^N \nu_m^{(0)}(x_j)\right] |x_1,\ldots,x_N\rangle. \quad (22)$$

For the standard set of ultra-local conservation laws we have[10]

$$\nu_m^{(0)}(x) = J\left(\frac{\sin\gamma}{\gamma}\frac{\partial}{\partial x}\right)^{m-1} \frac{\sin^2\gamma}{\cos\gamma - \cosh\gamma x}. \quad (23)$$

Minimizing the generalized Gibbs free energy[7] of the truncated GGE[6]

$$G_{n_0} = \sum_{n=1}^{n_0} \lambda_n H^{(n)} - S, \quad (24)$$

where $S$ is the entropy, leads to the generalized TBA equations

$$\epsilon_j(x) = \epsilon_j^{(0)}(x) + \sum_{k=1}^{\ell} \text{sgn}(q_k) T_{jk} * \ln\left[1 + e^{-\epsilon_k}\right]\bigg|_x ,$$

$$a_j(x) = \text{sgn}(q_j)[\rho_j^p(x) + \rho_j^h(x)] + \sum_{k=1}^{\ell} T_{jk} * \rho_k^p\bigg|_x , \quad (25)$$

where now

$$\epsilon_j^{(0)}(x) = \sum_{n=1}^{n_0} \lambda_n \nu_{n,j}^{(0)}(x),$$

$$\nu_{n,j}^{(0)}(x) = \left(\frac{\sin\gamma}{\gamma}\frac{\partial}{\partial x}\right)^{n-1} \epsilon_j^{(0)}(x), \quad (26)$$

and the relation to particle and hole densities is

$$\frac{\rho_j^h(x)}{\rho_j^p(x)} = e^{\epsilon_j(x)}. \quad (27)$$

The generalized TBA equations (25) have a form that is similar to the finite temperature ones (17). However, the "driving terms" $\epsilon_j^{(0)}(\lambda)$ are different in the two cases, and this leads to significant differences in the solutions $\epsilon_j(x)$ in the two cases.

Eqns (25) characterize the saddle point state of the density matrix (1). In order to obtain the correct representative state for our given initial conditions, we have to fix the Lagrange multipliers $\lambda_j$ such that the constraints (2) are fulfilled. Given a set of distribution functions for particles, the expectation values of the conservation laws can be calculated

$$i_m = \lim_{L\to\infty} \frac{\langle I_m \rangle}{L} = \sum_{j=1}^{\ell} \int dx\, \nu_{m,j}^{(0)}(x)\, \rho_j^p(x). \quad (28)$$

In practice we characterizing the initial density matrix by computing $\{i_m\}$ numerically, and we are therefore restricted to retaining only very few (one or two) higher conservation laws. Given the expectation values $\{i_m\}$ we then solve the generalized TBA equations (25) under the constraints

$$i_m - \sum_{j=1}^{\ell} \int dx\, \nu_{m,j}^{(0)}(x)\, \rho_j^p(x) = 0, \quad (29)$$

self-consistently by iteration.

### D. "Excitations" over the GGE equilibrium state

Once we have obtained a solution $\{\rho_j^{p,h}(x)\}$, we may construct a particular eigenstate ("representative state"[9]) of the Hamiltonian that gives rise to these densities in the thermodynamic limit, i.e. go to an appropriate microcanonical ensemble. This eigenstate is characterized by a solution of the DTE (11) for a particular set of (half-odd) integer numbers $I_\alpha^j$

$$L\theta_j(x_\alpha^j) - \sum_{k=1}^{\ell}\sum_{\beta=1}^{M_k} \Theta_{j,k}(x_\alpha^j - x_\beta^k) = 2\pi I_\alpha^j. \quad (30)$$

Energy and momentum of this state are given by (13)

$$E_0 = \sum_{j=1}^{\ell}\sum_{\alpha=1}^{M_j} \epsilon_j^{(0)}(x_\alpha^j)\,, \quad P_0 = \sum_{j=1}^{\ell}\sum_{\alpha=1}^{M_j} \frac{2\pi I_\alpha^j}{L}. \quad (31)$$

We may now construct "excitations" over this eigenstate by standard techniques. Small changes in the $I_\alpha^j$ give rise to solutions $\{\tilde{x}_\alpha^j\}$ of the DTE, which are very similar to the solution $\{x_\alpha^j\}$ for the representative state. For our purposes it is sufficient to consider a single particle-hole excitation for $m$-strings, because energies and momenta are additive. This corresponds to a solution where all (half-odd) integers are the same as for the representative state, except for $m$-strings, where one of the (half-odd) integers, denoted by $I_h^m$, is replaced by $I_p^m$. The corresponding DTE read

$$L\theta_j(\tilde{x}_\alpha^j) - \sum_{k=1}^{\ell}\sum_{\beta=1}^{M_k} \Theta_{j,k}(\tilde{x}_\alpha^j - \tilde{x}_\beta^k)$$
$$= \Theta_{j,m}(\tilde{x}_\alpha^j - x_p^m) - \Theta_{j,m}(\tilde{x}_\alpha^j - x_h^m) + 2\pi I_\alpha^j. \quad (32)$$

In addition there are the equations determining the precise positions $x_{p,h}^m$ of the particle and the hole, but we do not need them here. Energy and momentum of this state are given by (13)

$$E = \sum_{j=1}^{\ell}\sum_{\alpha=1}^{M_j} \epsilon_j^{(0)}(\tilde{x}_\alpha^j)\,, \quad P = P_0 + I_p^m - I_h^m. \quad (33)$$

In order to work out the differences $\Delta E(x_p^m, x_h^m) = E - E_0$ and $\Delta P(x_p^m, x_h^m) = P - P_0$ it is useful to introduce shift-functions[3]

$$F_j(x_\alpha^j) \equiv \frac{x_\alpha^j - \tilde{x}_\alpha^j}{x_{\alpha+1}^j - \tilde{x}_\alpha^j}. \quad (34)$$

Taking the difference between (30) and (32) and then going over to the thermodynamic limit we obtain a set of *linear* integral equations for the shift functions

$$F_j(x)\text{sgn}(q_j)\left[1 + e^{\epsilon_j(x)}\right] + \sum_k T_{jk} * F_k\bigg|_x$$
$$= \frac{1}{2\pi}\left[\Theta_{j,m}(x - x_p^m) - \Theta_{j,m}(x - x_h^m)\right]. \quad (35)$$

A useful short-hand notation for this system of equations is

$$F_j(x) - \sum_k K_{jk} * F_k\bigg|_{x_j} = f_j^{(0)}(x), \quad (36)$$

where we have defined

$$f_j^{(0)}(x) = \frac{\text{sgn}(q_j)}{2\pi} \frac{\Theta_{j,m}(x - x_p^m) - \Theta_{j,m}(x - x_h^m)}{1 + e^{\epsilon_j(x)}},$$

$$K_{jk}(x, y) = -\text{sgn}(q_j) \frac{T_{jk}(x - y)}{1 + e^{\epsilon_j(x)}}. \quad (37)$$

We may write down a formal solution of the set of integral equations by inverting (36)

$$F_j(x) = (I - K)_{jk}^{-1} * f_k^{(0)} \Big|_x. \quad (38)$$

The "excitation energy" is then

$$\Delta E(x_p^m, x_h^m) = \epsilon_m^{(0)}(x_p^m) - \epsilon_m^{(0)}(x_h^m)$$
$$+ \sum_k \int dx_k \, (\epsilon_k^{(0)}(x_k))' F_k(x_k). \quad (39)$$

The second line is rewritten as

$$\sum_k \int dx_k \, (\epsilon_k^{(0)}(x_k))' (I - K)_{kj}^{-1} * f_j^{(0)} \Big|_{x_k}. \quad (40)$$

We now define functions $e_j'(x)$ by

$$e_j' * (I - K)_{jk} \Big|_{x_k} = (\epsilon_k^{(0)}(x_k))'. \quad (41)$$

Inverting the integral operator gives

$$e_j'(x_j) = (\epsilon_m^{(0)})' * (I - K)_{mj} \Big|_{x_j}, \quad (42)$$

which in turn allows us to recast expression (39) for the energy difference in the form

$$\Delta E(x_p^m, x_h^m) = \epsilon_m^{(0)}(x_p^m) - \epsilon_m^{(0)}(x_h^m)$$
$$+ \int dx_j \, e_j'(x_j) f_j^{(0)}(x_j). \quad (43)$$

Finally, using that

$$\frac{\partial f_j^{(0)}(x_j)}{\partial x_p^m} = K_{jm}(x_j, x_p^m), \quad (44)$$

we obtain

$$\frac{\partial \Delta E(x_p^m, x_h^m)}{\partial x_p^m} = e_m'(x_p^m). \quad (45)$$

This allows us to express the energy in the form

$$\Delta E(x_p^m, x_h^m) = e_m(x_p^m) - e_m(x_h^m). \quad (46)$$

The momentum of the excitation is simply

$$\Delta P(x_p^m, x_h^m) = \frac{2\pi}{L}(I_m^p - I_m^h) = z_m(x_p^m) - z_m(x_h^m), \quad (47)$$

where $z_m(x)$ is the counting function for $m$-strings. The counting functions can be obtained from a set of coupled, nonlinear integral equations, but for our purposes it suffices that

$$\frac{dz_j(x)}{dx} = 2\pi \left[\rho^p(x) + \rho^h(x)\right]. \quad (48)$$

The conclusion of these considerations is that the particle contributes $e_m(x_p^m)$ to the energy and $z_m(x_p^m)$ to the momentum. Furthermore, contributions from particles and holes are additive. The group velocity of the particle is then given by

$$v_m(x) = \frac{\partial e_m(x)}{\partial z_m(x)} = \frac{\frac{\partial e_m(x)}{\partial x}}{\frac{\partial z_m(x)}{\partial x}}$$
$$= \frac{e_m'(x)}{2\pi \rho_m^p(x) \left(1 + e^{\epsilon_m(x)}\right)}. \quad (49)$$

In order to determine the velocities of the various elementary excitations, we first solve the gTBA equations as described above, substitute the solution for the dressed energies into the integral equations (41) for the functions $e_j'(x)$, and finally solve the latter numerically by iteration.

## II. FINITE-TEMPERAUTRE TIME EVALUATING USING METTS

The idea of minimally entangled typical thermal states, put forward by White[11], is to devise a Markov process to importance sample wave functions such that *i)* the ensemble average of the expectation values of some operator equals the expectation value with respect to a thermal (Gibbs) ensemble and *ii)* the entanglement entropy of the sampled wave functions is small such that they can be represented efficiently via matrix product states. We will briefly review the algorithm which is detailed in Refs. 11,12 and comment on the implementation of the time evolution.

To be specific, consider the expectation value of some operator with respect to the thermal density matrix $\rho_\beta = 1/Z \exp(-\beta H)$, where $Z = \text{tr}[\exp(-\beta H)]$ is the partition function and $\beta = 1/(k_\text{B} T)$ is the inverse temperature. It reads

$$\langle \hat{O} \rangle_\beta = \text{tr}[\hat{O} \rho_\beta] = \frac{1}{Z} \sum_{\vec{i}} \langle \vec{i} | e^{-\beta H/2} \hat{O} e^{-\beta H/2} | \vec{i} \rangle \quad (50)$$

and can be expanded in an orthonormal basis $\{|\vec{i}\rangle\}$. One then defines a set of properly normalized states via the mapping

$$|\Psi_{\vec{i}}\rangle = \frac{e^{-\beta H/2}|\vec{i}\rangle}{\sqrt{\langle \vec{i}|e^{-\beta H}|\vec{i}\rangle}} \quad (51)$$

such that equation (50) is cast into the form

$$\langle \hat{O} \rangle_\beta = \sum_{\vec{i}} P(\vec{i}) \langle \Psi_{\vec{i}} | \hat{O} | \Psi_{\vec{i}} \rangle \qquad (52)$$

where $P(\vec{i}) = \langle \vec{i} | e^{-\beta H} | \vec{i} \rangle$ can be considered as the "Boltzmann weight" of state $|\vec{i}\rangle$.

The idea of importance sampling is to generate states $|\Psi_{\vec{i}}\rangle$ with probability $P(\vec{i})$ such that the weighted sum in equation (52) can be approximated by an ensemble average over states

$$\langle \hat{O} \rangle_\beta \approx \overline{\langle \Psi_{\vec{i}} | \hat{O} | \Psi_{\vec{i}} \rangle}. \qquad (53)$$

The METTS algorithm is then a Markov chain whose stationary distribution samples the states $|\Psi_{\vec{i}}\rangle$ with the correct distribution and the procedure works as follows[11,12]:

1. Start from a random classical product state (CPS) $|\vec{i}\rangle$.

2. Evolve the CPS in imaginary time, $|\psi_{\vec{i}}\rangle = e^{-\beta H/2} |\vec{i}\rangle$

3. Project back into a CPS basis with probability $p(i \to j) = \langle \psi_{\vec{i}} | \vec{j} \rangle$.

4. Continue with 2.

Eventually, the states $|\psi_{\vec{i}}\rangle$ will thermalize towards the correct METTS ensemble $|\Psi_{\vec{i}}\rangle$.

A couple of remarks are in order. First, the autocorrelation time of the Markov chain strongly depends on the CPS basis. In particular, the naive approach only projecting into the $S_z$ basis (which would allow the efficient use of magnetization as a conserved quantity) is found to converge only extremely slowly. Projection into different bases such as alternating between the $S_z$ and $S_x$ basis restores ergodicity and helps the algorithm to properly explore the full phase space which results in a very fast thermalization[11]. Second, the projection in step 3. can be done without calculating all overlaps (which would take exponentially long) but can be implemented using a successive projection using matrix product operators as detailed in Ref. 12.

To implement the quench we first generate the ensemble of METTS by starting from random CPS and applying the previously outlined projection procedure. For each state, we use a statistically independent random CPS initial state to avoid autocorrelation effects. The expectation values of the operators after the quench are then evaluated via the operator $\hat{O}(t)$ evolved in time with respect to the final Hamiltonian $H(\Delta_f)$. Inserting this into equation (53) yields

$$\langle \hat{O}(t) \rangle_\beta \approx \overline{\langle \Psi_{\vec{i}} | e^{iH(\Delta_f)t} \hat{O} e^{-iH(\Delta_f)t} | \Psi_{\vec{i}} \rangle}. \qquad (54)$$

This means that we have evolve each state of our METTS ensemble in time (we do so by using the standard TEBD[13] algorithm) and are then able to calculate ensemble averages over the evolved ensemble at each time step.

* Electronic address: lars.bonnes@uibk.ac.at
[1] M. Fagotti and F.H.L. Essler, J. Stat. Mech. P07012 (2013); M. Fagotti, M. Collura, F.H.L. Essler and P. Calabrese, Phys. Rev. B **89**, 125101 (2014).
[2] M. Rigol, V. Dunjko, V. Yurovsky, and M. Olshanii, Phys. Rev. Lett. **98**, 50405 (2007).
[3] V.E. Korepin, A.G. Izergin, and N.M. Bogoliubov, *Quantum Inverse Scattering Method, Correlation Functions and Algebraic Bethe Ansatz* Cambridge University Press, 1993;
[4] M. P. Grabowski and P. Mathieu, Annals Phys. 243, 299 (1995).
[5] T. Prosen and E. Ilievski, Phys. Rev. Lett. **111**, 057203 (2013).
[6] M. Fagotti and F.H.L. Essler, Phys. Rev. B **87**, 245107 (2013).
[7] J.-S. Caux and J. Mossel, J. Phys. A: Math. Theor. 45, 255001 (2012).
[8] M. Takahashi, *Thermodynamics of One-Dimensional Solvable Models*, Cambridge University Press, Cambridge 1999.
[9] J.-S. Caux and F.H.L. Essler, Phys. Rev. Lett. **110**, 257203 (2013).
[10] N. Muramoto and M. Takahashi, J. Phys. Soc. Jpn **68**, 2098 (1999).
[11] S. R. White, Phys. Rev. Lett. **102**, 190601 (2009)
[12] E. M. Stoudenmire and S. R. White, New. J. Phys. **12**, 055026 (2010).
[13] G. Vidal, Phys. Rev. Lett. **91**, 147902 (2003)

 

\end{document}